\documentclass[useAMS,usenatbib]{mnras}
\usepackage[latin1]{inputenc}
\usepackage[english]{babel}
\usepackage{graphicx}
\usepackage{amsmath}
\usepackage{amsfonts}
\title[On Generalized Theories of Varying Fine Structure Constant]{On Generalized Theories of Varying Fine Structure Constant}
\author[Soumya Chakrabarti]{Soumya Chakrabarti\thanks{E-mail : soumya1989@bose.res.in}\\
Department of Theoretical Sciences\\
S. N. Bose National Centre for Basic Sciences\\
Kolkata, West Bengal 700106\\
India}
\date{Accepted XXX. Received YYY; in original form ZZZ}
\pubyear{2022}
\begin{document}
\maketitle
\begin{abstract}
We work with a class of scalar extended theory of gravity that can drive the present cosmic acceleration as well as accommodate a mild cosmic variation of the fine structure constant $\alpha$. The motivation comes from a vintage theory developed by Bekenstein, Sandvik, Barrow and Magueijo. The $\alpha$ variation is introduced by a real scalar field interacting with charged matter. We execute a cosmological reconstruction based on a parametrization of the present matter density of the Universe. Observational consistency is ensured by comparing the theoretical estimates with JLA + OHD + BAO data sets, using a Markov chain Monte Carlo simulation. An analysis of molecular absorption lines from HIRES and UVES spectrographs is considered as a reference for the variation of $\alpha$ at different redshifts. Two examples are discussed. The first explores a field-dependent kinetic coupling of the scalar field interacting with charged matter. The second example is a generalized Brans-Dicke formalism where the varying $\alpha$ is fitted in as an effective matter field. This generates a simultaneous variation of the Newtonian constant $G$ and $\alpha$. The pattern of this variation can have a crucial role in cosmic expansion history.    
\end{abstract}
\section{Introduction}\label{s0}
In physics, \textit{`Nature'} is usually defined as a coalescence of phenomenology on different energy scales. For this notion to uphold, one requires simple and mathematically consistent theories. More often than not, these theories need to introduce new structures, as in fields or symmetries or more importantly, fundamental constants. The title `fundamental constant' is allotted only to a few pre-assigned parameters which can not be derived. They can only be measured and other parameters of the theory can be expressed in terms of them. These constants define patches of the physical world in an axiomatic manner while their origin remains a riddle. Any idea that accommodates a possible variation of these constants is a trial of the standard theories and a motivation to think beyond our usual understanding of physical reality. \\

The \textit{`Large Numbers hypothesis'} of \cite{dirac1, dirac2} can be thought of as a precursor to the idea of a varying fundamental constant. The hypothesis tips that the universal constants can be different for different phases of the evolving universe and should be treated as varying entities. Most of the early attempts (see for instance the review of \cite{unzicker}) to implement this notion into a theory of gravity went astray, until a successful field-theoretic approach was considered by \cite{jordan}, allowing variations of gravitational coupling as well as the fine-structure constant. The observational viability of this theory was subsequently discussed by \cite{fierz}. A special example was considered soon after by \cite{bransdicke} with only the gravitational coupling being varied. This particular case is now popularly known as the Brans-Dicke theory and regarded as the pioneer of a larger class of theories called the Scalar-Tensor theories \citep{damour1}. The possibility of a cosmic variation of fine structure constant $\alpha$ at Hubble rate was first proposed by \cite{gamow}. This idea boosted interest in a cosmological theory with varying $\alpha$ immediately, leading to a series of works in succession challenging Gamow's original claim. For instance, nuclear mass systematics \citep{peres, dyson1, dyson2} and the laboratory analysis of the fine-structure splittings in radiogalaxy emission lines \citep{bahcall} did predict a much more mildly varying $\alpha$ than the Hubble rate. Comparison of cesium atomic clocks and superconducting cavity clocks \citep{turne}, observation of active galactic nucleus such as a BL-Lacertae object \citep{wolfe} and the analysis of fission product isotopes in natural reactors also require particular attention \citep{akhter} in this regard. More advanced constraints on this variation can be found from the \textit{$SZ$ to $X$-ray flux ratio analysis} of galaxy clusters \cite{desai}, and from the relativistic transitions in molecular absorption lines of Quasar spectra at different redshifts \citep{savedoff, bahcall1, nunes, parkinson, doran, webbetal, murphy, uzan1, chand}. The latter will be of particular interest in this manuscript. More inclusive and motivating discussions on theories allowing a variation of natural constants are available in the literature \citep{uzan1, uzan2, chiba} and may provide further insights to the readers. \\

We focus on a theory of gravity that allows a mild variation of $\alpha$ in the cosmological past and an asymptotic approach to the desired value at present. Theories of unification (e.g. string theory) inspire such a cosmic variation by arguing that a coupling in $4$-dimension is just a projection of fundamental constants defined in higher dimensions \citep{antoniadis}. A General Relativistic (GR) description of this variation primarily requires a dynamical framework and an evolution equation for $\alpha$. A rather radical approach of constructing this is through a variation of the speed of light $c$ where $\alpha = e^{2}/c\hbar$ \citep{moffat, albrecht}. Although theories with varying $c$ have received some interest in possible resolutions of cosmological problems \citep{barrow1, barrow2}, they come along with a general breakdown of Lorentz invariance. We work with a class of theory where the electron charge is written as an evolving real scalar field (the $e$-field) which drives a variation of $\alpha$. The philosophy goes in parallel with standard GR where the Hubble function is driven by the total energy density of the universe through the cosmological equations; and here the evolving density of an electromagnetic entity does the trick. We receive motivation from the theory of gravity proposed by \cite{bekenstein}. The original theory is simple and consistent about standard requirements such as general covariance, Lorentz invariance, causality, and scale invariance of the $e$-field. There is a general breakdown of charge conservation but the theory remains a nice platform to combine GR with Maxwell's theory of electromagnetism \citep{landau}. The theory needs to be generalized to satisfy cosmological requirements \citep{barrow3, magukibble}. In particular, the class of generalized theory developed by \cite{bsbm} is now popular as the Bekenstein-Sandvik-Barrow-Magueijo or the \textit{BSBM} theory. We aim to work with extensions of a standard BSBM theory and the resulting non-trivial variations in space-time geometry, especially from a cosmological purview. \\

In many ways, GR has left behind some riddles in cosmology that are difficult to decode. For example, the standard theory and most of its existing modifications fail to allow a smooth transition of the universe into the present acceleration from a preceding era of deceleration. This is now confirmed by advanced astrophysical observations, such as the Luminosity Distance measurement of Supernova \citep{riess1, betoule}. Theoretically, the effective fluid distribution of Dark Energy (DE) driving this phenomenon should have an evolving Equation of State (EOS) \citep{maor1, maor2, upadhye}. It is easier to illustrate this using the Deceleration parameter $q$ which should evolve into negative values somewhere in the recent past \citep{paddy1, paddy2}. This is exactly where approaches like energy corrections of the order of a cosmological constant goes awry; in reference with the contradictions with observations \citep{riess2, eisenstein} and the \textit{`coincidence problem'} \citep{ralf, velten}. Simple scalar fields are used to develop toy models which can drive different phases of the cosmic expansion depending on whether the scalar self-interaction dominates over the kinetic part or not \citep{zlatev, sahni, copeland}. Any scalar extended theory of gravity carries some motivations from theories of unification but all are not equally acceptable. For example, models that introduce a dark energy through slow-rolling scalar fields can not account for a variation in the EOS \citep{sensensami, slepian} and are effectively ruled out. Similarly, the Quintessence models become superfluous in view of the constraints on scalar-baryon interaction, derived from the \textit{`fifth-force experiments'} \citep{adel}. A trick perhaps lies in the choice of the self-interaction, for example, a quantum field theory inspired pseudo-Nambu-Goldstone-Boson (pNGB) case \citep{frieman} works remarkably well in cosmological aspects. An interesting alternative is to allow the scalar field to interact with ordinary matter such that it decouples in high density regions of the cosmos (e.g. around the earth) and avoids detection from local experiments \citep{khoury, hinter}. These generalizations give a set of constraints such that the theory does not allow a violation of the Equivalence Principle (EP) on the solar system scales \citep{jain, will0, will, gubser, upadhye1, brax, damourpoly}. However, recent research on the screening of Milky Way galaxy predicts that a scalar field satisfying the solar system constraints can not, on their own, drive the present acceleration of the universe \citep{wang}. Therefore, it is natural to keep looking for a better theoretical framework that can support the cosmic acceleration without violating basic observational requirements. \\

An extended BSBM-type theory of gravity can potentially provide this framework, fitting in beautifully with a number of requirements. The scalar field in this theory is, in fact, a prototype of the so-called chameleon fields and readily deals with questions regarding possible equivalence principle violations. A few examples of extended BSBM have already received attention in cosmological context \citep{barrowprd, barrowplb, barrowlip}. We work with two different generalizations of BSBM. The first one allows the coupling constant in the kinetic term governing the $e$-field dynamics to be a function of the field itself. The second example is a unification of an extended BSBM and a generalized Brans-Dicke (BD) theory where the fine structure constant $\alpha$ and the Newtonian constant $G$ can vary simultaneously. We take a field-dependent kinetic coupling of the $e$-field and assume the BD parameter $\omega_{BD}$ to be a function of the BD scalar. It is well known that the standard BD setup, while widely acknowledged, could not quite deliver a \textit{`better theory of gravity'}. This is primarily due to the observational constraints on $\omega_{BD}$ \citep{bertotti} and the lack of viable cosmological solution for all epochs \citep{banerjeesen, faraoni}. A generalization like field-dependent $\omega_{BD}$ can potentially resolve these issues and provide a more complete formulation of the theory. We refrain ourselves from choosing any functional form of the field profiles or their interactions. We formulate a simple way to reverse engineer the structure of the theory, a cosmological reconstruction from $Om(z)$ parametrization. This parameter has received quite rigorous attention in the analysis of observational data and subsequent comparisons of modified dark energy models \citep{sahni1, sahni2, lu, tong}. $Om(z)$ is a constant $\sim\Omega_{m0}$ for standard $\Lambda$CDM model, denoting the present matter density of the universe. The reconstruction gives an observationally viable Hubble function as a function of redshift which is used to solve the field equations of the generalized theories. We discuss the role of different components of the theory in different epochs of the cosmic evolution and try to identify which components can play a probable key role in driving the transition between successive epochs. The evolution of the $e$-field with redshift tells us how the fine structure constant might have gone through a mild evolution alongside the cosmic expansion. In addition, the second generalization gives us a scope to study the simultaneous evolution of $G$ and $\alpha$ with redshift. We also try to provide an idea how these fundamental couplings can be directly related to one another, plotting $\alpha$ as a function of $G$. This inter-relation is the hint of a more general background formalism (perhaps geometric) relating all the fundamental couplings, which is at this moment beyond our scope.  \\

In Section $2$, we briefly review the standard BSBM formalism. In Section $3$, we discuss a bit about the observational constraints to be considered throughout the manuscript. This includes an execution outline of the reconstruction based on $Om(z)$ and the analysis of data from Molecular Absorption Spectroscopy. Sections $4$ and $5$ include discussions on the structure of generalized BSBM theories. We make some concluding remarks and finish in Section $6$.

\section{The Bekenstein-Sandvik-Barrow-Magueijo (BSBM) theory : A brief review}
We have already mentioned that in the standard BSBM theory, the electron charge $e$ is assumed to be a function of coordinates. It is written as a dimensionless scalar field $\epsilon$ called the $e$-field. 
\begin{equation}
e = e_{0}\epsilon (x^\mu).
\end{equation}
$e_{0}$ has the dimension of $e$. All particle charges vary uniformly through this universal scalar.
\begin{equation}
\alpha = e^{2}/c\hbar,
\end{equation}
where $c$ and $\hbar$ are constants. \medskip

A constant $\alpha$ is a signature of Maxwell's electrodynamics. Here, the vector potential interacts minimally with matter which works as a \textit{`dictum'} to help us decide which adjustments in the standard laws of physics are rational and which are not. A variation of $\alpha$ requires some careful tweaking of the standard Maxwellian construct, for example, by allowing a general breakdown of charge conservation. Moreover, the modified theory must allow a model-independent framework and abide by generally acknowledged principles of physics. For instance, we should be able to derive the dynamical equations for $\alpha$ from an invariant action using a corresponding action principle. We must also have second-order, hyperbolic evolution equations to ward off non-causality or runaway solutions. \\ 

The Lorentz invariant Lagrangian for a charged particle in flat spacetime is written as
\begin{equation}\label{actionreview}
L = -mc(-u^{\mu}u_{\mu})^{\frac{1}{2}} + \frac{e_{0}\epsilon}{c}u^{\mu}A_{\mu},
\end{equation}
where $m$ is the rest mass and $e_{0}\epsilon$ is the charge of the particle. The proper time is written as $\tau$ and the four-velocity as $u^{\mu} = \frac{dx^{\mu}}{d\tau}$. We note that the Lagrangian has a minimally interacting vector potential term and is invariant under the gauge transformation 
\begin{equation}
\epsilon A_\mu = \epsilon A_\mu + \chi_{,\mu},
\end{equation}
$\chi$ being an arbitrary function. The electromagnetic field is identified from Eq. (\ref{actionreview}) by writing the Lagrange equation
\begin{equation}\label{lagrangereview}
\frac{d}{d\tau}\left[m u_{\mu} + \frac{e_{0}}{c}\epsilon A_{\mu} \right] = -m_{,\mu}c^{2} + \frac{e_{0}}{c}(\epsilon A_{\nu})_{,\mu}u^{\nu}.
\end{equation}
$u_{\mu}u^{\mu} = -c^{2}$ is the normalization used above and the rest mass is written as a function of coordinates. Eq. (\ref{lagrangereview}) can be simplified into
\begin{equation}\label{lagrangereview2}
\frac{d(m u_{\mu})}{d\tau} = -m_{,\mu}c^{2} + \frac{e_{0}}{c} \left[ (\epsilon A_{\nu})_{,\mu} - A_{\mu})_{,\nu} \right] u^{\nu}.
\end{equation}
Eq. (\ref{lagrangereview2}) is important for two reasons. Primarily, it introduces the concept of anomalous force through the term $m_{,\mu}c^{2}$ \citep{dicke}. More importantly, one can identify a gauge-invariant electromagnetic field from the Lorentz force term (second term on the RHS) and write, following \cite{barrowprd}
\begin{equation}
F_{\mu\nu} = \frac{\left\lbrace (\epsilon A_\nu)_{,\mu} - (\epsilon A_\mu)_{,\nu}\right\rbrace}{\epsilon}.
\end{equation}
$F_{\mu\nu}$ is also invariant under a constant rescaling of $\epsilon$. The corresponding lagrangian contribution is written as
\begin{equation}
{\cal L}_{em} = -F^{\mu\nu}F_{\mu\nu}/4.
\end{equation}
However, an evolution equation for $\epsilon$ can not be written from this simple construct. A separate lagrangian which can govern the dynamics of $\epsilon$ and satisfy the dimensional requirement, was given by Bekenstein \citep{bekenstein} as
\begin{equation}
{\cal L}_\epsilon = -\frac{1}{2}\omega \frac{(\epsilon_{,\mu }\epsilon^{,\mu})}{{\epsilon^2}}.
\end{equation}

For a standard BSBM theory $\omega = \frac{\hbar c}{l^2}$. $l$ comes in as a dimensional correction and works as a length scale of the theory. The theory allows the electric field to be Coulombic for a point charge only above this length scale. This puts additional constraints on the corresponding energy scale $\frac{\hbar c}{l}$. A generalization of this setup was given by \cite{bsbm} using a transformed gauge
\begin{equation}
a_\mu =\epsilon A_\mu,
\end{equation}
and a modifed field tensor
\begin{equation}
f_{\mu\nu} = \epsilon F_{\mu\nu} = \partial_\mu a_\nu - \partial_\nu a_\mu.
\end{equation}

Using $\psi = ln\epsilon$ as a variable, the action is written as
\begin{equation}\label{standbsbn}
S = \int d^4x \sqrt{-g}\left({\cal L}_g+{\cal L}_{mat}+{\cal L}_\psi +{\cal L}_{em}e^{-2\psi }\right).
\end{equation}

The setup has a similarity with dilaton-type theories \citep{forgacs, marciano, barrowsingle, damourpoly}. While a standard dilaton field interacts with standard matter, $\psi$ or the $e$-field in BSBM interacts only with the electromagnetic sector. In Eq. (\ref{standbsbn}) the $\psi$-contribution is
\begin{equation}
{\cal L}_\psi = -{\frac{\omega}{2}}\partial_\mu \psi \partial^\mu \psi,
\end{equation}
and the electromagnetic contribution is
\begin{equation}
{\cal L}_{em}=-\frac{1}{4}f_{\mu \nu }f^{\mu \nu}.
\end{equation}

$\omega$ is a coupling constant and ${\cal L}_g = \frac{1}{{16\pi G}}R$ is the standard GR part. A usual metric variation of the action produces the modified field equations
\begin{equation}
G_{\mu \nu }=8\pi G\left( T_{\mu \nu }^{mat} + T_{\mu \nu }^\psi +T_{\mu \nu}^{em}e^{-2\psi }\right).
\end{equation}

Similarly, a $\psi$ variation produces the scalar field evolution equation responsible for $\alpha$-dynamics
\begin{equation}\label{boxeq}
\Box \psi = \frac 2\omega e^{-2\psi }{\cal L}_{em}. 
\end{equation}
 
Before moving forward to the next section, let us write the cosmological equations for a standard BSBM setup. The independent field equations for a spatially-flat FRW geometry are 
\begin{eqnarray}
\left(\frac{\dot{a}}a\right)^2 = \frac{8\pi }3\left( \rho _m\left( 1+\zeta
_m e^{-2\psi}\right) +\rho_{r}e^{-2\psi} + \frac{\omega}{2}\dot{\psi}^2 \right),
\label{fried1}
\end{eqnarray}
and the $\psi$ evolution equation
\begin{equation}
\ddot{\psi} + 3H\dot{\psi} = -\frac{2}{\omega} e^{-2\psi}\zeta_{m}\rho_{m}.
\label{psiddot}
\end{equation}

All the equations are in $G = c = 1$ unit. One needs to specify the nature of $L_{em}$ to write these equations. This is done from a parametrization $\zeta = \frac{L_{em}}{\rho}$. $\rho$ is the total baryon energy density. This parameter defines the non-relativistic matter contribution in $L_{em}$. $\zeta_{m}$ is the cosmological value of $\zeta$ which depends on the non-baryonic matter content of the universe, including Dark matter. The interplay between electric and magnetic fields in the cold dark matter distribution is tipped to be a crucial factor that restricts the cosmological value of $\zeta$ to be between $-1$ and $+1$ \citep{barrowprd}. Based on comparison with spectroscopic analysis of molecular absorption lines from Quasars, a model with negative $\zeta_{m}$ is thought to be slightly disfavored \citep{barrowlip}. However, it is more rational to call these arguments speculative since there exists, to date, no clear knowledge of a Dark matter distribution. The fluid contents of the effective energy-momentum distribution, i.e., radiation and the matter, generate their respective conservation equations
\begin{eqnarray}
&&\dot{\rho_m}+3H\rho_m = 0, \\&& \label{dotrhom}
\dot{\rho_r} + 4H\rho_r = 2\dot{\psi}\rho_r.  \label{dotrho}
\end{eqnarray}

A solution of this set of Eqs. (\ref{fried1}), (\ref{psiddot}), (\ref{dotrhom}) and (\ref{dotrho}) dictates the cosmic evolution of fine structure constant, written as
\begin{equation}
\alpha =\exp (2\psi )e_{0}^2/\hbar c.
\end{equation}
However, an exact solution in analytical form is never guaranteed. It is tactically better to work with some ansatz over one of the components of the equations, preferably supported by pheneomenological evidences. This is where a scheme of reconstruction can prove to be much more effective. In the next section we discuss about the particular scheme we are interested in before moving on to generalized BSBM setups.

\section{Comparison with Observational Data}
\subsection{Cosmological Reconstruction from {\it $Om(z)$}}
Reconstruction is a way to develop a theory in reverse order. One usually starts from one or more widely accepted phenomenological facts, directly or indirectly inspired by astrophysical observations. For example, in cosmology, a natural intuition would be to start from an optimum behavior any parameter that governs the evolution of the universe. The field equations of the theory can then be solved to write the best possible structure. Reconstruction schemes based on dimensionless kinematic quantities are particularly popular \citep{bern, visser, catto, duna}. These quantities are written as parameters involving higher-order derivatives of the scale factor, such as deceleration, jerk or statefinder \citep{alam, mukherjee, chakrabarti1}. In this manuscript, we work with the $Om(z)$ parameter, which has gained much popularity for its utility in categorical comparison of cosmological models with observational data. The parameter is a measure of the present matter density contrast of the universe, written as a constant value $\Omega_{m0}$ for standard cosmology. Quite a number of schemes are popular based on this parameter \citep{sahni1, sahni2, lu, tong}. We illustrate a rather simple method to parametrize $\Omega_{m0}$, introducing two new parameters at the outset. This allows a mild variation of the parameter with redshift. The aim is to compare this parametric form with astrophysical observations and to estimate the new parameters for a best possible behavior for Hubble and deceleration. The standard parameter is written as

\begin{eqnarray}
&& \Omega_{m0} = \frac{h(z)^{2}-1}{(1+z)^{3}-1}, \\&&
h(z) = \frac{H(z)}{H_{0}}.
\end{eqnarray}
The present value of Hubble is given by $H_{0}$. The parametrization is introduced by replacing $\Omega_{m0}$ with

\begin{equation}\label{ansatz}
Om(z) = \lambda_{0} (1 + z)^{\delta}.
\end{equation}

Alternatively, we can write Hubble as a function of redshift
\begin{equation}\label{hubbleansatz}
h(z) = \left[1 + \lambda_{0} (1+z)^{\delta} \left\lbrace (1+z)^{3} - 1 \right\rbrace\right]^{\frac{1}{2}},
\end{equation}

in closed analytical form. We estimate the model parameters using a statistical analysis, in comparison with data-sets provided by observations from : (i) the Joint Light Curve Analysis of Supernova distance modulus ($SDSS-II$ and $SNLS$ collaborations) \citep{betoule}, (ii) the Hubble parameter measurements (OHD) \citep{jimenez, stern, blake, moresco, chuang, planck, delubac} and (iii) the Baryon Acoustic Oscillation (BAO) data ($6dF$ $Galaxy$ $Survey$, BOSS LOWZ and BOSS CMASS) \citep{beutler, boss}. The argument of differentiation in Eq. (\ref{hubbleansatz}) is changed into redshift $z$ and the Hubble is written in its dimensionless form by scaling $H_0$ by $100$ km $\mbox{Mpc}^{-1}$ $\mbox{sec}^{-1}$
\begin{equation}
h(z) = \frac{H(z)}{H_0} = \frac{H(z)}{100\times h_0}.
\end{equation}

\begin{figure}
\begin{center}
\includegraphics[angle=0, width=0.52\textwidth]{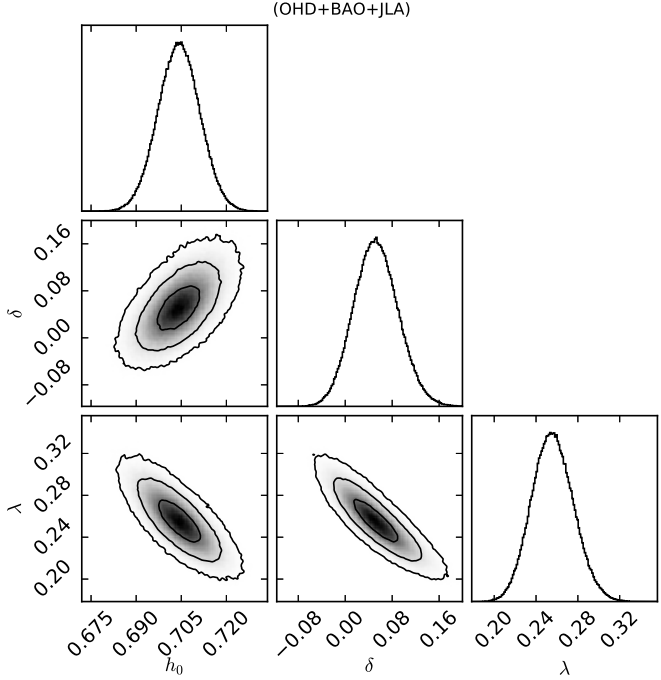}
\caption{Best fit parameter and uncertainty estimation (1$\sigma$ and 2$\sigma$) using confidence contours using a combined OHD+JLA+BAO data set.}
\label{Modelcontour}
\end{center}
\end{figure}

We estimate the present value of dimensionless Hubble ($h_{0} = H_{0}/100 km\mbox{Mpc}^{-1} \mbox{sec}^{-1}$) and the deceleration parameter directly using a \textit{Markov Chain Monte Carlo} simulation (MCMC) written in python \citep{pythonmcmc}. The parameter space confidence contours reveal the best fit values alongwith associated uncertainty in the estimation, as shown in Fig. \ref{Modelcontour}. We also write the best-fit parameter values and 1$\sigma$ error estimation in Table. \ref{resulttable} for convenience. \\

\begin{table}
\caption{{\small Best Fit measurement of present values of dimensionless Hubble, $\delta$ and $\lambda_{0}$ with 1$\sigma$ uncertainty in estimation.}}\label{resulttable}
\begin{tabular*}{\columnwidth}{@{\extracolsep{\fill}}lrrrrl@{}}
\hline
 & \multicolumn{1}{c}{$h_0$} & \multicolumn{1}{c}{$\delta$} & \multicolumn{1}{c}{$\lambda_{0}$} \\
\hline
$OHD+JLA+BAO$ 	  & $0.704^{+0.007}_{-0.007}$ &$0.052^{+0.038}_{-0.037}$ & $0.256^{+0.020}_{-0.019}$ &\\
\hline
\end{tabular*}
\end{table}

\begin{figure}
\begin{center}
\includegraphics[angle=0, width=0.40\textwidth]{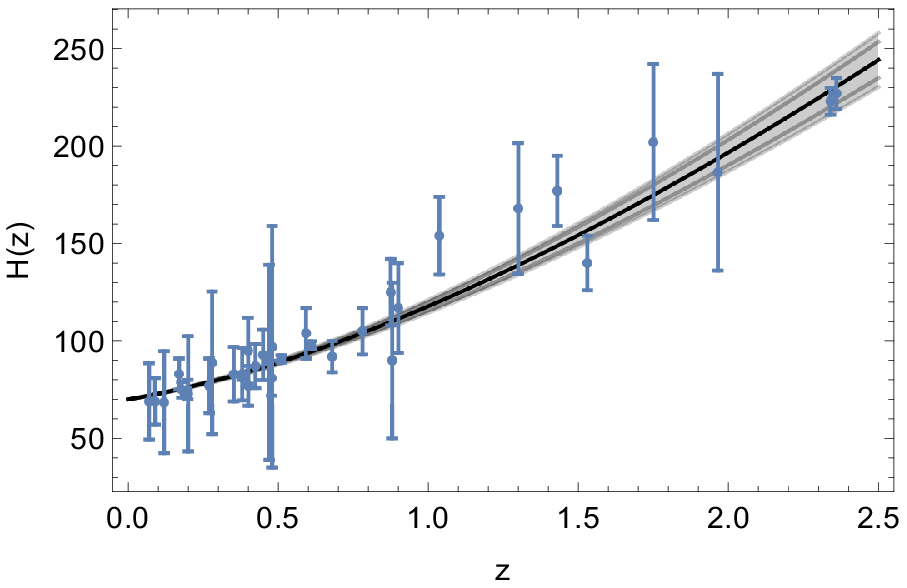}
\includegraphics[angle=0, width=0.40\textwidth]{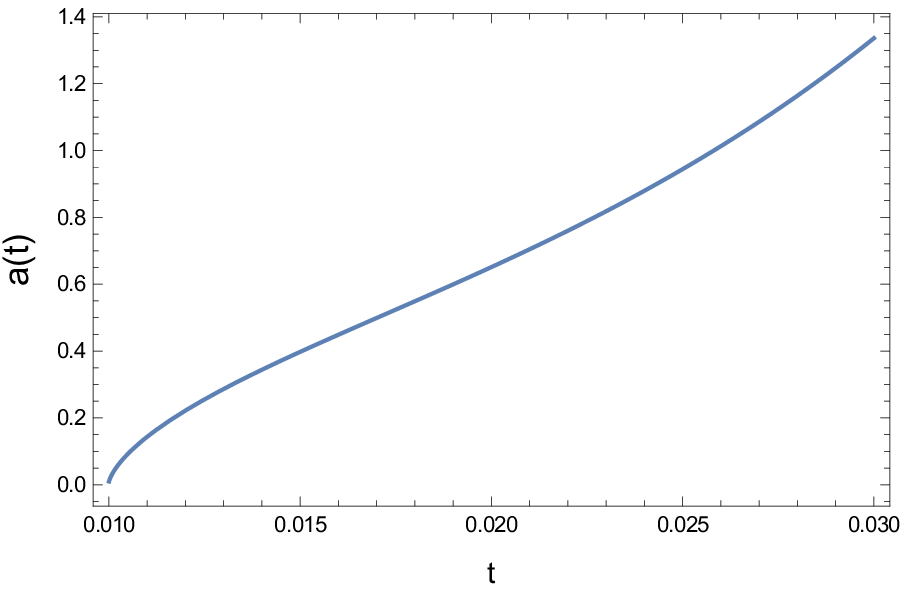}
\caption{Top Panel : Evolution of $H(z)$ with redshift fitted against observation, for the best fit parameter values of $H_0$, $\delta$ and $\lambda_{0}$ (thick black line) and for 2$\sigma$, 3$\sigma$ region of uncertainty (gray shaded regions). Bottom Panel : Scale factor as a function of cosmic time showing a smooth succession of acceleration after deceleration.}
\label{Hz_data}
\end{center}
\end{figure}

The present value of Hubble parameter is well consistent with recent observations \citep{planck}. A departure from standard cosmology can be noted, particularly from the estimated best fit value of $\delta$. While for a $\Lambda$CDM model $\delta$ would be exactly $0$, for the extended cosmology the parameter is found to be in the range $0.052^{+0.038}_{-0.037}$. There is also a slight difference in the estimate of present matter density contrast, given by $\lambda{0}$. These departures can be used to identify modified models of Dark Energy and categorize their behavior as Quintessence-like or Phantom-like \citep{shafieloo, wangteg}. However, an advanced cosmological analysis is not within the scope of this manuscript and we are happy with an overall optimum behavior of the evolving DE. The Hubble $H(z)$ is plotted in Fig. \ref{Hz_data} alongwith the data points from observation. This evolution has sufficient observational validity for a reasonable range of redshift. To demonstrate the transition into present acceleration from a deceleration we also plot the numerical solution of scale factor with cosmic time, for a convenient range of reference. \\

\begin{figure}
\begin{center}
\includegraphics[angle=0, width=0.40\textwidth]{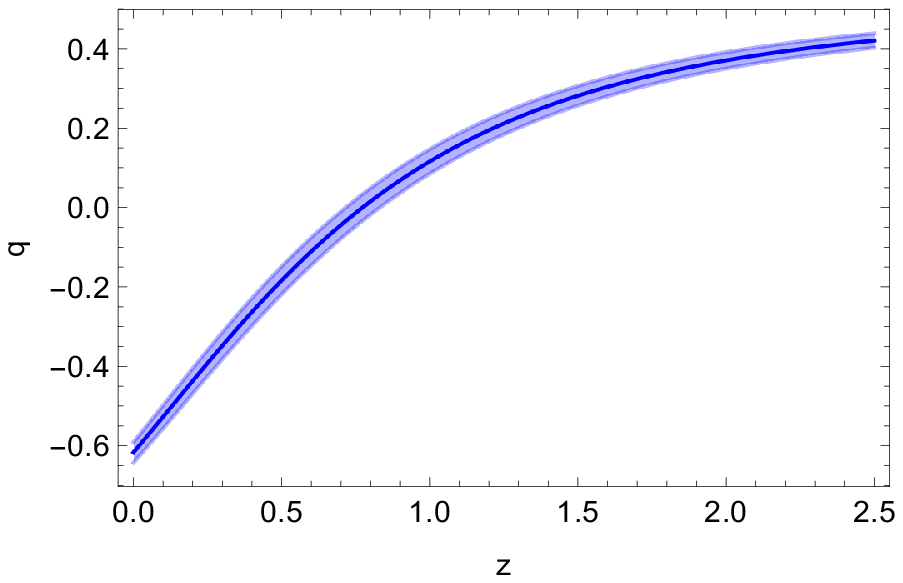}
\includegraphics[angle=0, width=0.40\textwidth]{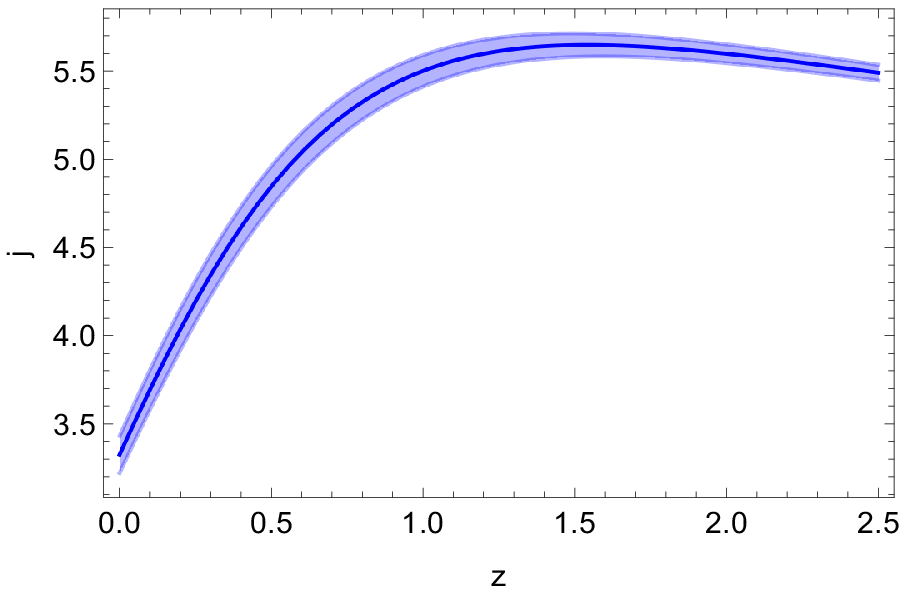}
\caption{Top Panel : Deceleration parameter as a function of z. Bottom Panel : Jerk parameter as a function of z. The plots are for the best fit parameter values of $H_0$, $\delta$ and $\lambda_{0}$ (bold blue) as well as for the regions of uncertainty (faded blue shades).}
\label{cosmic_parameters}
\end{center}
\end{figure}

This transition can be better understood from the acceleration term $\ddot{a}$. This is explained in Fig. \ref{cosmic_parameters} where we plot $q(z)$ and the jerk parameter $j(z)$, both for the best fit parameters (bold blue) and for the regions of uncertainty (faded blue shades). The present value of $q(z)$ is found to be $\sim -0.62$. The deceleration moves from a positive zone into negative at a redshift $z_{t} < 1$, which marks the transition of the universe. This transition redshift as well as the present value of deceleration provide a good agreement with observational requirements. We also plot the jerk parameter in the bottom panel of Fig. \ref{cosmic_parameters} whose evolution confirms that higher-order parameters show a clear departure compared to a $\Lambda$CDM (for which $j = 1$) in this extended cosmology. \\

We note at this point that the reconstructed expansion rate and dynamical evolutions are independent of the theory of gravity under consideration. The total energy-momentum distribution of the universe has an effective EOS written as
\begin{equation}
w_{eff} = \frac{p_{tot}}{\rho_{tot}}.
\end{equation}

Writing the present critical density as $\rho_{c0} = 3H^{2}_0/8\pi G$, the effective EOS can be written as a function of the expansion rate using
\begin{eqnarray}
&& \frac{\rho_{tot}}{\rho_{c0}} = \frac{H^2(z)}{H^2_0}, \\&&
\frac{p_{tot}}{\rho_{c0}} = -\frac{H^2(z)}{H^2_0}+\frac{2}{3}\frac{(1+z)H(z)H'(z)}{H^2_0}.
\end{eqnarray}

\begin{figure}
\begin{center}
\includegraphics[angle=0, width=0.40\textwidth]{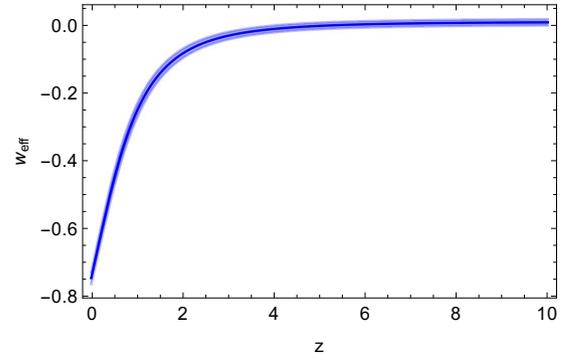}
\caption{The effective equation of state as a function of $z$. The plot is for the best fit parameter values of $H_0$, $\delta$ and $\lambda_{0}$ (bold blue) as well as for the regions of uncertainty (faded blue shades).}
\label{weff}
\end{center}
\end{figure}

In Fig. \ref{weff} we plot $w_{eff}$ as a function of redshift. It is negative at $z\sim 0$ with a present value close to $-1$. This confirms an effective negative pressure during the DE dominated late-time acceleration. For larger redshifts, $w_{eff}$ approaches zero suggesting the earlier deceleration of the universe being matter-dominated.  

Apart from this, we expect the growth of matter over-density to follow a corresponding $\Lambda$CDM pattern closely. This is necessary to avoid a large departure from the observed structure formation in the universe. We look into this by first assuming that the background density is homogeneous, written as $\rho_m$. Small deviations from $\rho_m$ are written as $\delta\rho_m$. The `matter density contrast' is then defined as
\begin{equation}
\delta_m = \frac{\delta\rho_m}{\rho_m}.
\end{equation}

$\delta_m$ can not accelerate away with Hubble or the scale factor. Its dynamics is governed by a non-linear evolution equation in locally over-dense distributions such as around a star or a collapsing distribution. However, for a spatially homogeneous late-time cosmology a linearized evolution equation is efficient enough.
\begin{equation}\label{delm_eq}
\ddot{\delta}_m + 2H\dot{\delta}_m = 4\pi G\rho_m\delta_m.
\end{equation}

\begin{figure}
\begin{center}
\includegraphics[angle=0, width=0.40\textwidth]{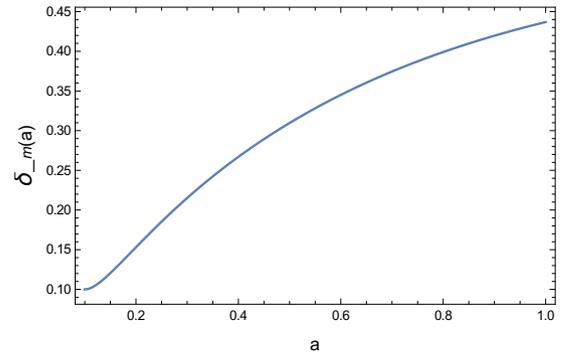}
\caption{Plot of $\delta_m$ vs $a$ for the best fit parameter values of $H_0$, $\delta$ and $\lambda_{0}$.}
\label{overdensity}
\end{center}
\end{figure}

We change the time derivatives of Eq. (\ref{delm_eq}) into derivatives with respect to scale factor and solve for $\delta_m$ numerically. We take the initial conditions for scale factor as $a_i = 0.01$ and for overdensity as $\delta_m(a_i) = 0.01$ and $\dot{\delta}_m(a_i) = 0$. The solution for the best fit parameter values is plotted in Fig. \ref{overdensity} and a behavior closely following $\Lambda$CDM model can be seen. \\

Before concluding this section we discuss the thermodynamic equilibrium of the cosmological system in brief. The notion is that a thermodynamic system consisting of the universe surrounded by a cosmological horizon is not too different from a black hole in thermodynamic equilibrium \citep{gibbons, jacobson, paddythermo}. To recover the first law of blackhole thermodynamics for a spatially flat cosmological system the horizon is given by $r_h = 1/H$ and is known as the Hubble horizon \citep{bakrey}. The total entropy of this system surrounded by the Hubble horizon must not decrease with cosmic expansion. We write the total entropy $S$ as a sum of the boundary entropy and the entropy of the constituent fluid

\begin{equation}
S = S_h + S_f.
\end{equation}

The constraints on the total entropy are written as
\begin{eqnarray}\label{thermoreq}
&& \frac{dS}{dn} \geq 0, \\&&
\frac{d^2S}{dn^2} < 0. \\&&\nonumber
n=\ln{a}.
\end{eqnarray}
The horizon entropy is proportional to the horizon area ${\mathcal A} = 4 \pi {r_h}^{2}$. In a $\hbar = k_B = c = 8\pi G = 1$ unit it is written as, 
\begin{equation}
S_h = 8\pi^2 {r_h}^{2}.
\end{equation}

The temperature of the horizon depends on its radius \citep{jacobson, bakrey, frolov}
\begin{equation}
T_h = 1/2\pi r_{h}.
\end{equation}

Assuming that a late time cosmology is dominated by a coexistence of dark energy and dark matter, the total fluid entropy is 
\begin{equation}
S_f = S_{cdm} + S_{de}.
\end{equation}

Then the first law of thermodynamics gives
\begin{eqnarray}
&& TdS_{cdm}=dE_{cdm}+p_{m}dV=dE_{cdm}, \\&&
TdS_{de}=dE_{de}+p_{d}dV.
\end{eqnarray}
 
$p_d$ and $p_m$ are the pressures of two fluid components and the fluid temperature $T$ is uniform. If $V = 4\pi {r_h}^{3}/3$, the energy contributions are written as
\begin{eqnarray}
&& E_{cdm}=\frac{4\pi r^3_h\rho_{cdm}}{3}, \\&&
E_{de}=\frac{4\pi r^3_h\rho_{de}}{3}.
\end{eqnarray}

If we assume that $T = T_{h}$, i.e., the horizon has the same temperature as the fluid, the rate of change of entropy with time can be written as

\begin{equation}
\dot{S}=\dot{S}_{cdm}+\dot{S}_{de}+\dot{S}_{h}=4\pi Hr_h^2[\rho_m+(1+w_{de})\rho_{de}]^2.
\label{Sdot}
\end{equation}

Eq. (\ref{thermoreq}) ensures a thermodynamic equilibrium depending on how the total entropy $S$ changes with $n = \ln{a}$ in the first order and in the second order. Using Eq. (\ref{Sdot}), we can write 
\begin{eqnarray}\label{Sdn}
&& S_{,n} = \frac{16\pi^2}{H^4}(H_{,n})^2, \\&&
S_{,nn} = 2S_{,n}\left(\frac{H_{,nn}}{H_{,n}}-\frac{2H_{,n}}{H}\right) = 2S_{n}\Psi.
\end{eqnarray}

\begin{figure}
\begin{center}
\includegraphics[width=0.40\textwidth]{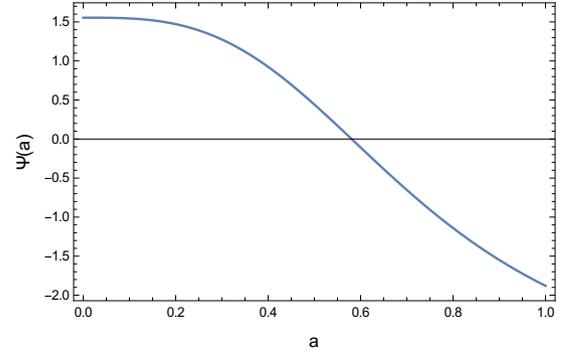}
\caption{Evolution of $\Psi = \left(\frac{H_{,nn}}{H_{,n}}-\frac{2H_{,n}}{H}\right)$ with respect to scale factor for the best fit parameter values of $H_0$, $\delta$ and $\lambda_{0}$.}
\label{psiplot}
\end{center}
\end{figure}

We note that $\Psi < 0$ ensures the thermodynamic equilibrium. For the present work, all we need to do is to bring in the exact form of Hubble in Eq. (\ref{Sdn}) from Eq. (\ref{hubbleansatz}) and plot $\Psi$ as a function of the scale factor $a$. Fig. \ref{psiplot} shows the plot and one can find $\Psi$ to be negative during late-times. There is an interesting evolution from positive into the negative domain during the cosmic expansion which hints that perhaps the universe tends to move towards a thermodynamic equilibrium. We should mention here that $\dot{S}$ may have some additional role in the estimation of cosmological quantities for different types of dark energy models, as discussed quite recently by \cite{jamil}. 

In a nutshell, this simple formulation is used to write a desired late-time cosmological behavior based on a list of observations. It can not be denied that this is restricted only in the late-time era. For a unified picture of $\alpha$ variation along with cosmic expansion, one requires a different analysis altogether. Moreover, this is not the best possible structure, but it can provide an overall qualitative idea going by usual conventions. However, with the closed form of Hubble we will be able to avoid any speculative assumptions to simplify the field equations. Now, in subsequent sections, we will demonstrate the application of this scheme for two extended versions of BSBM theory. In particular, we need to solve for the $e$-field and determine the evolution of the fine structure coupling. For that, we also need to fit in with measurements of Quasar absorption spectra which is briefly discussed in the next subsection. 

\subsection{Observational Constraints of $\alpha$}
In the introductory notes, it is already mentioned that mild cosmic variation of different fundamental couplings are observed and reported in the literature \citep{uzan2, martins}. They receive serious attention particularly in view of the standard model of particle physics. Most of these observations come from relativistic transitions in molecular absorption lines of Quasar spectra at different redshifts. These provide enough motivation for one to keep looking beyond standard cosmology. Even if the considerations are somewhat unorthodox, they can potentially provide practical solutions to avoid violations of the Equivalence Principle \citep{martins1, leite, webb1}. In this work we refer to a combined analysis of constraints from molecular absorption spectroscopy \citep{webb1, ferreira1, ferreira2, whitmore, pinho, martinspinho} and the late-time cosmological observations.  \\

\begin{table}
\begin{center}
\begin{tabular}{|c|c|c|c|}
\hline
 Source & Redshift & ${\Delta\alpha}/{\alpha}$ (ppm) & Spectrograph. \\
\hline\hline
J0026$-$2857 & 1.02 & $3.5\pm8.9$ & UVES \\
\hline
J0058$+$0041 & 1.07 & $-1.4\pm7.2$ & HIRES \\
\hline
3 sources & 1.08 & $4.3\pm3.4$ & HIRES \\
\hline
HS1549$+$1919 & 1.14 & $-7.5\pm5.5$ & UVES/HIRES/HDS \\
\hline
HE0515$-$4414 & 1.15 & $-1.4\pm0.9$ & UVES \\
\hline
J1237$+$0106 & 1.31 & $-4.5\pm8.7$ & HIRES \\
\hline
HS1549$+$1919 & 1.34 & $-0.7\pm6.6$ & UVES/HIRES/HDS \\
\hline
J0841$+$0312 & 1.34 & $3.0\pm4.0$ & HIRES \\
J0841$+$0312 & 1.34 & $5.7\pm4.7$ & UVES \\
\hline
J0108$-$0037 & 1.37 & $-8.4\pm7.3$ & UVES \\
\hline
HE0001$-$2340 & 1.58 & $-1.5\pm2.6$ &  UVES \\
\hline
J1029$+$1039 & 1.62 & $-1.7\pm10.1$ & HIRES  \\
\hline
HE1104$-$1805 & 1.66 & $-4.7\pm5.3$ & HIRES \\
\hline
HE2217$-$2818 & 1.69 & $1.3\pm2.6$ &  UVES \\
\hline
HS1946$+$7658 & 1.74 & $-7.9\pm6.2$ & HIRES \\
\hline
HS1549$+$1919 & 1.80 & $-6.4\pm7.2$ & UVES/HIRES/HDS \\
\hline
Q1103$-$2645 & 1.84 & $3.5\pm2.5$ &  UVES \\
\hline
Q2206$-$1958 & 1.92 & $-4.6\pm6.4$ &  UVES \\
\hline
Q1755$+$57 & 1.97 & $4.7\pm4.7$ & HIRES  \\
\hline
PHL957 & 2.31 & $-0.7\pm6.8$ & HIRES  \\
PHL957 & 2.31 & $-0.2\pm12.9$ & UVES  \\
\hline
\end{tabular}
\caption{\label{table2} The dataset for $\alpha$ variations, written as ${\Delta\alpha}/{\alpha}$ in a unit of parts per million. The absorption spectra are for different Quasar sources, measured by different spectrographs at different redshifts. Some values are written as weighted average of separate independent measurements \citep{songaila}.}
\end{center}
\end{table}

Generally, variations of three entities, proton-to-electron mass ratio $\mu$, the fine structure constant $\alpha$ and the proton gyromagnetic ratio $g_p$ are reported in a combined form \citep{webb1, ferreira1, ferreira2, whitmore}. The observed variations are indeed quite mild, with the scale of $\sim 10^{-16} \, year^{-1}$. For mathematical convenience they are written in comparison with Hubble constant $H_{0} \simeq 7 \times 10^{-11} \, year^{-1}$ in a scale $\simeq 10^{-6} H_{0}$ or parts per million (ppm). We take a specific set of measurements for $(\alpha_{z}-\alpha_{0}) / \alpha_{0} = \Delta \alpha / \alpha$ from HIRES and UVES spectrographs \citep{reim, ferreira3, songaila, evans, kotus, agafonova, molaro}, respectively at the Keck and VLT telescopes. The tightest available measurements for each source are written in Table. \ref{table2}. We compare them with the evolution of $\Delta\alpha/\alpha$ derived by solving the modified field equations.

\section{Generalized BSBM : Theory and Reconstruction}
We first apply this analysis to a simple extension of BSBM theory that was introduced by \cite{barrowlip}. This extension allows $\omega$, the coupling of the kinetic term to be a function of the $e$-field, $\psi$. We briefly discuss the action and the field equations of the theory without making any particular choice of $\omega(\psi)$. The reconstruction allows us more freedom to solve the field equations numerically and see how $\psi$ might have evolved in the recent past. From this, we can determine an optimum dynamics for $\alpha$ that does not alter the course of a viable late-time cosmology. More importantly, we can compare the theoretical evolution of $\frac{\Delta \alpha}{\alpha}$ with the data from molecular absorption spectra as in Table. \ref{table2}. \\

We work with the Lagrangian 
\begin{equation}
\mathcal{L} = \sqrt{-g}(\mathcal{L}_{g} + \mathcal{L}_{\text{mat}} + \mathcal{L}_{\psi } + \mathcal{L}_{\text{em}}e^{-2\psi}).  \label{LAG}
\end{equation}
Apart from the usual gravitational part 
\begin{equation}
\mathcal{L}_{g} = \frac{R}{16\pi G},
\end{equation}
there is a contribution of the scalar field 
\begin{equation}
\mathcal{L}_{\psi } = -\frac{\omega(\psi)}{2}\partial_{\mu}\psi \partial ^{\mu}\psi,
\end{equation}
and the electromagnetic part 
\begin{equation}
\mathcal{L}_{\text{em}} = -\frac{1}{4}f_{\mu\nu }f^{\mu\nu},
\end{equation}
in the Lagrangian. The condition $\omega(\psi) \geqslant 0$ is known as the \textit{no-ghost} condition and is enforced at the outset for a positive energy density of the $e$-field. $\alpha$ evolves as 
\begin{equation}
\alpha =\alpha _{0} e^{2\psi},  \label{alpha}
\end{equation}
similar to a standard BSBM setup. In natural units, the spatially-flat FRW equation with a $-,+,+,+$ signature are written from a metric variation of Eq. (\ref{LAG}) 
\begin{equation}
\frac{\dot{a}^{2}}{a^{2}} = \frac{8\pi}{3}\left(\rho _{m}\left(1+|\zeta|e^{-2\psi}\right) + \rho_{nb} + \rho_{r}e^{-2\psi} + \rho_{\psi}\right).   \label{FRIED}
\end{equation}

The derivatives are for comoving proper time and written as overdots. The energy-momentum content on the RHS consists of four parts. Contribution from standard matter and non-baryonic matter are written as $\rho_{m}$ and $\rho_{nb}$. Both of them are proportional to $a^{-3}$, however, $\rho_{nb}$ is supposed to fit in for a cold dark matter component and therefore written as a separate entity. The electromagnetic part of the field equation is written using the ratio $\zeta = \frac{L_{em}}{\rho}$. The ratio defines the non-relativistic matter contribution to $L_{em}$ which can be better understood in reference with Eq. (\ref{boxeq}). The scalar field contribution to energy density is 
\begin{equation}
\rho_{\psi} = \frac{\omega(\psi)}{2}\dot{\psi}^{2}.
\end{equation}

We have not included any separate constant energy density correction such as a $\rho_{\Lambda}$, as we want to see the $e$-field profile that can drive a late-time cosmology and at the same time maintain the desired mild variation of $\alpha$, all by itself. Its evolution is governed by
\begin{equation}
\ddot{\psi}+3H\dot{\psi}+\frac{\omega ^{\prime}(\psi)}{2\omega }\dot{\psi}^{2}=-\frac{2\zeta \rho _{m}}{\omega}e^{-2\psi}.  \label{COSM}
\end{equation}

The noninteracting radiation density $\rho_r$ is covariantly conserved, leading to the condition
\begin{equation}\label{radiationdef}
\rho_{ir} = \rho_{r}e^{-2\psi} \propto a^{4}.
\end{equation}

As discussed by \cite{barrowlip}, a GR limit of the theory can be defined by the conditions $|\zeta|e^{-2\psi }\ll 1$ and $\dot{\psi}^{2}\omega \ll \rho_{c}$ under which Eq. (\ref{FRIED}) gives back the standard Friedmann equation. $\rho_{c}$ is the primary matter content of the epoch one is considering, i.e., different for a cold dark matter or a $\Lambda$ epoch. We do not enforce these conditions at the outset as our aim is a bit different, to formulate a scalar dominated theory exhibiting enough departure from GR. However, we recall that the interaction of electric and magnetic fields in a cold dark matter distribution is expected to constrain the cosmological value of $\zeta$ heavily \citep{barrowprd} and therefore, we do take $\zeta$ to be between $-1$ and $+1$. \\

To solve Eqs. (\ref{FRIED}), (\ref{COSM}) and (\ref{radiationdef}) we call in the reconstruction of Hubble from $Om(z)$ parametrization, given by Eqs. (\ref{ansatz}) and (\ref{hubbleansatz}). We take the best fit parameter values of $\lambda_{0}$, $\delta$ and $H_{0}$ from Table \ref{resulttable}, and use
\begin{equation}
H(z) = H_{0}\left[1 + \lambda_{0} (1+z)^{\delta} \left\lbrace (1+z)^{3} - 1 \right\rbrace\right]^{\frac{1}{2}}.
\end{equation}
First, we change the arguments of the equations from cosmic time into redshift and change the derivatives. To demonstrate in brief,

\begin{eqnarray}\label{1eq}
&& \dot{\psi} = \frac{d\psi}{dz} \frac{dz}{da}\frac{da}{dt}, \\&&\nonumber
= -H_{0}{\psi^\circ}(1+z)\left[1 + \lambda_{0} (1+z)^{\delta} \left\lbrace (1+z)^{3} - 1 \right\rbrace\right]^{\frac{1}{2}}, \\&&
\end{eqnarray}
where we have used $\psi^\circ = d\psi/dz$ and $a = 1/(1+z)$. Using this the second derivative of $\psi$ can be written as

\begin{align}\label{2eq}\nonumber
\ddot{\psi} &= H_{0}^{2}(1+z)^{2} {\psi^{\circ\circ}}\Big[ 1 + \lambda_{0}(1+z)^{\delta}\Big\lbrace (1+z)^3 - 1\Big\rbrace \Big] \\\nonumber
&\quad + H_{0}^{2}(1+z){\psi^\circ} \Big[ 1 + \lambda_{0}(1+z)^{\delta}\Big\lbrace (1+z)^3 - 1\Big\rbrace \Big] \\\nonumber
&\quad +\frac{H_{0}^{2}}{2}(1+z)^{2}{\psi^\circ} \Big[\delta \lambda_{0}(1+z)^{\delta - 1}\Big\lbrace (1+z)^{3} - 1\Big\rbrace \\
&\quad + 3\lambda_{0}(1+z)^{\delta+2}\Big]\,.
\end{align}

We also use Eq. (\ref{FRIED}) to define $\omega(\psi)$ as

\begin{eqnarray}\label{omegaz}\nonumber
&& \omega(\psi) = \frac{2}{H_{0}^{2}{\psi^\circ}^{2}(1+z)^{2}\Big[1+\lambda(1+z)^{\delta}\Big\lbrace(1+z)^{3}-1\Big\rbrace\Big]}\\&&\nonumber
\Bigg[3H_{0}^{2}\Big[1+\lambda(1+z)^{\delta}\Big\lbrace(1+z)^{3}-1\Big\rbrace\Big] - \rho_{0}(1+\zeta e^{-2\psi})\\&&
(1+z)^{3} - \rho_{1}(1+z)^{4} - \rho_{2}(1+z)^{3}\Bigg].
\end{eqnarray}

\begin{figure}
\begin{center}
\includegraphics[angle=0, width=0.40\textwidth]{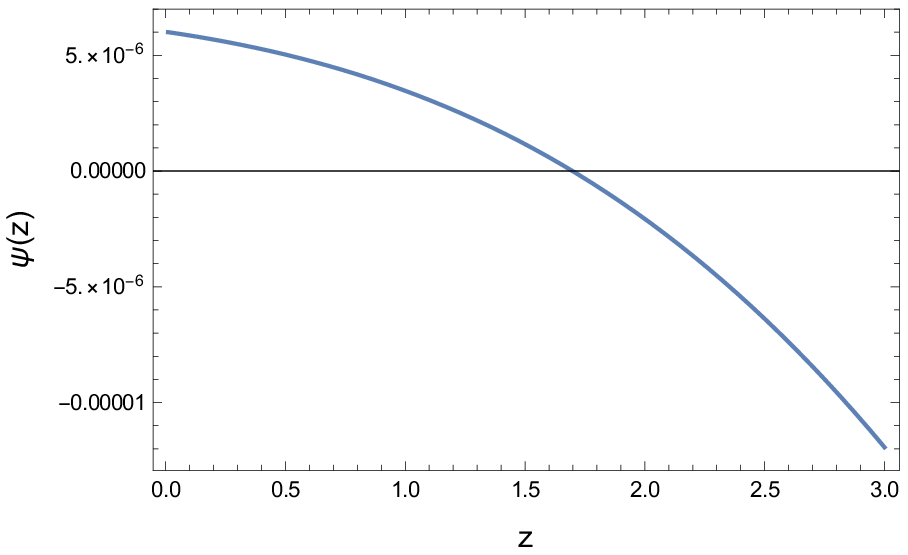}
\includegraphics[angle=0, width=0.40\textwidth]{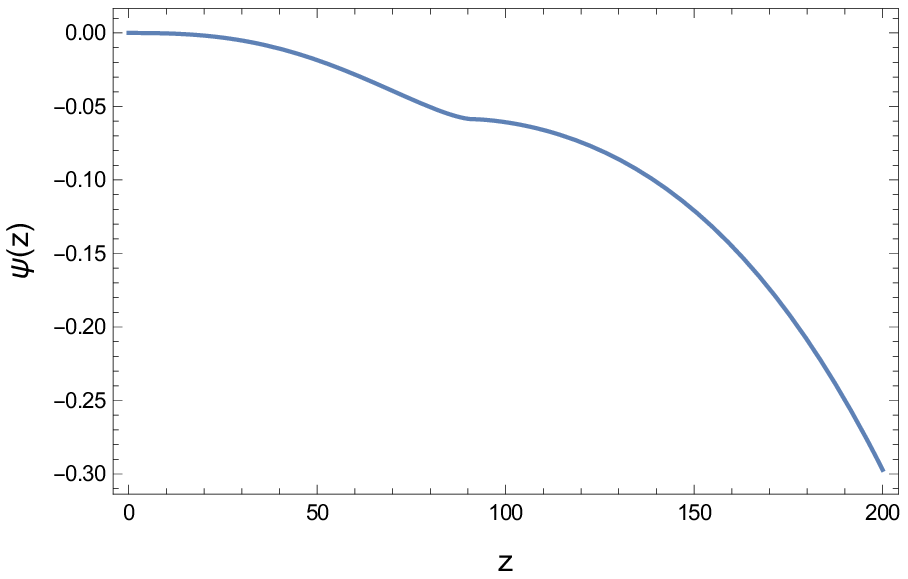}
\caption{Generalized BSBM : Evolution of $\psi$ in the low redshift regime (top panel) and high redshift regime (bottom panel) for the best fit parameter values of $\lambda_{0}$, $\delta$ and $H_{0}$.}
\label{psiscalar}
\end{center}
\end{figure}

$\rho_{0}$, $\rho_{1}$ and $\rho_{2}$ are proportionality constants. Using Eq. (\ref{omegaz}), we solve the $e$-field evolution Eq. (\ref{COSM}) numerically and plot $\psi$ as a function of redshift $z$ in the top panel of Fig. \ref{psiscalar}. It can be seen that just before the redshift of transition, $z_{t} \sim 1$, the scalar field crosses into a positive domain. For most of the earlier epochs, the scalar evolves through the negative sector. The overall variation of the scalar is mild $\sim \left[0.000005, -0.30\right]$. This can be understood from the bottom panel of Fig. \ref{psiscalar} where $\psi$ evolution is shown for larger redshifts as well. \\

\begin{figure}
\begin{center}
\includegraphics[angle=0, width=0.40\textwidth]{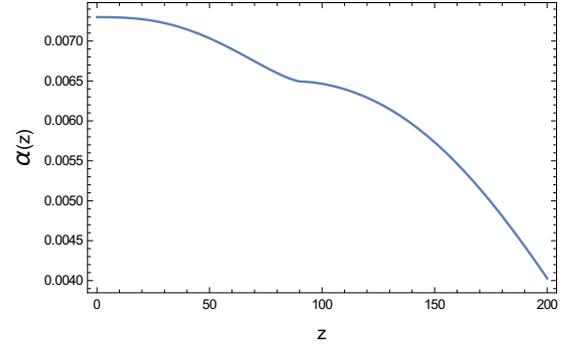}
\caption{Generalized BSBM : Evolution of $\alpha$ as a function of $z$ for the best fit parameter values of $\lambda_{0}$, $\delta$ and $H_{0}$.}
\label{alphaz}
\end{center}
\end{figure}
 
\begin{figure}
\begin{center}
\includegraphics[angle=0, width=0.40\textwidth]{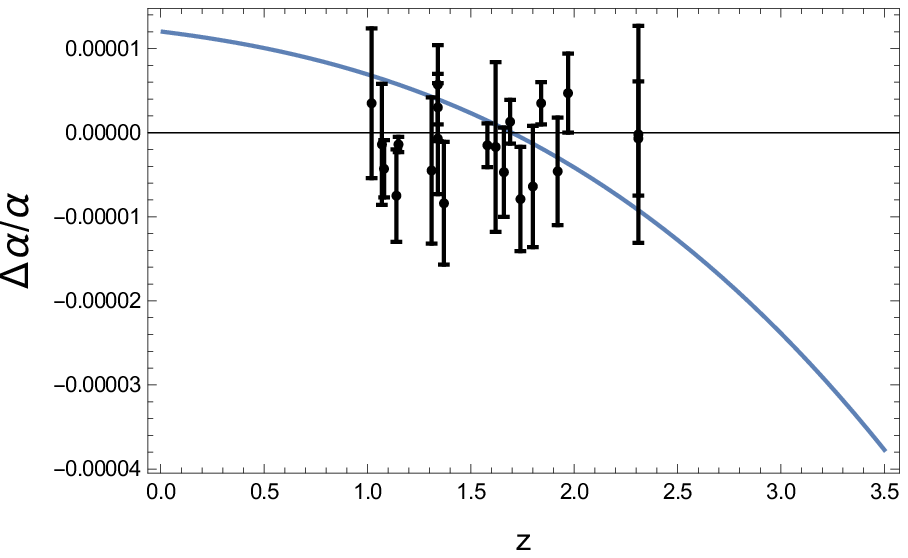}
\includegraphics[angle=0, width=0.40\textwidth]{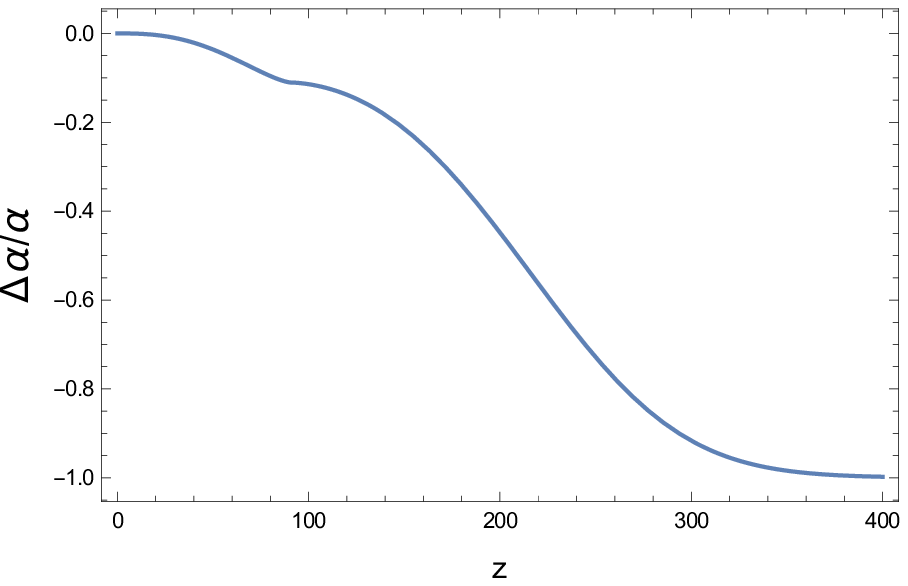}
\caption{Generalized BSBM : The top panel shows an evolving $\frac{\Delta \alpha}{\alpha}$ with $z$, fitted against a set of data from molecular absorption spectra. The bottom panel shows the evolution for larger redshifts. Both the plots are for the best fit parameter values of $\lambda_{0}$, $\delta$ and $H_{0}$.}
\label{deltaalpha}
\end{center}
\end{figure}

This already indicates a mild variation of the fine structure constant which is exponentially related to $\psi$. In Fig. \ref{alphaz} we plot $\alpha$ with $z$. Moreover, taking $\alpha_{0}$ and $\alpha_{z}$ to be the values of the coupling at the present epoch and at some redshift $z$, we derive the evolution of the quantity $(\alpha_{z}-\alpha_{0}) / \alpha_{0} = \Delta \alpha / \alpha$. The theoretically estimated $\Delta \alpha / \alpha$ is illustrated in Fig. \ref{deltaalpha} while the observational data points as in Table \ref{table2} are fitted in. The graph on the top panel shows that at low redshift, the theory gives a good fit with the observations of molecular absorption spectra. The bottom panel shows $\Delta \alpha / \alpha$ for a larger range of redshift, until $z \sim 400$. The variation is indeed mild even within this range, approaching a negligible variation for higher values of redshift. While this evolution is perfectly reasonable, it is primarily based on a reconstruction from present cosmological observations. Arguably, this may not be accurate enough for a larger redshift analysis.  \\
  
\begin{figure}
\begin{center}
\includegraphics[angle=0, width=0.40\textwidth]{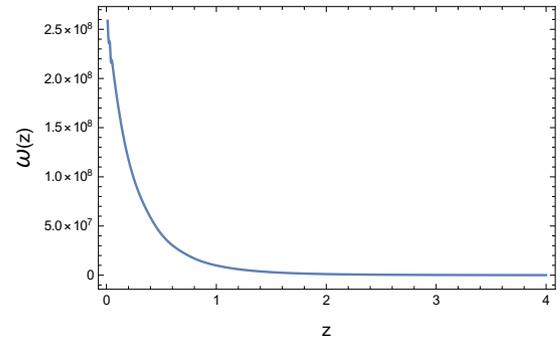}
\caption{Generalized BSBM : The evolving $\omega(\psi)$ as a function of $z$ for the best fit parameter values of $\lambda_{0}$, $\delta$ and $H_{0}$.}
\label{omegapsi}
\end{center}
\end{figure}

Using Eq. (\ref{omegaz}) we plot the evolution of $\omega(\psi)$ as a function of redshift in Fig. \ref{omegapsi}. The plot suggests that while $\omega(\psi)$ varies negligibly during deceleration, it starts increasing quite rapidly once the deceleration-into-acceleration transition sets in. This might play a crucial role in launching the onset of a late-time acceleration. As a general comment we can say that any accelerated regime under the scope of this theory is driven by a dominant kinetic contribution of the $e$-field. Since the reconstructed Hubble holds true over these numerical solutions of the field equations, the effective EOS of the system is effectively given in Fig. \ref{weff}. This means that the scalar field and its interaction with charged matter conspire with each other to create the present epoch of Dark Energy dominated acceleration ($w_{eff} \sim -1$ around $z \sim 0$) and a matter-dominated deceleration ($w_{eff} \sim 0$) in the recent past. \\

We must mention here that the numerical solutions show negligible qualitative differences when different signs of the parameter $\zeta$ are chosen, as far as the parameter is chosen within the range of $-1$ and $+1$. Moreover, one can easily extend this model to a further extent; for instance, by including a self-interaction $V(\psi)$ in the action. In such a case, keeping in mind that the scalar field itself evolves mildly, the potential $V(\psi)$ needs to be dominant enough in order to register any significant changes in the resulting dynamics. For instance, if one chooses $V(\psi) \sim V_{0} \psi^{n}$ or something like $V(\psi) \sim V_{0} \psi^{n} + V_{1} \psi^{m}$ or an exponential $V(\psi) \sim V_{0} + V_{1}e^{\sigma\psi}$, the parameters like $V_{0}$, $V_{1}$, $m$, $n$, $\sigma$ need to be large enough. However, with such a modification, $\alpha$ may evolve on unexpected scales and violate the observational constraints from molecular absorption spectroscopy.  \\
  
Before moving onto the next section we note that $\omega(\psi)$ takes quite a large value during the late-time acceleration compared to earlier epochs. The scale readily compels one to imagine this in comparison with $\omega_{BD}$, the standard Brans-Dicke parameter and the observations constraints on the same. We note that the scalar $\psi$ in this particular setup is not really a geometric scalar field as it interacts differently with the lagrangian. Therefore, driven by simple curiosity, considering a geometric field as well as an $e$-field in the Lagrangian seems to be a logical extension. We consider this in the next section.

\section{Generalized Brans-Dicke-BSBM : Theory and Reconstruction}
In this section, we work with a generalized Brans-Dicke-BSBM setup which can support a variation of the Newtonian coupling $G$ as well as the fine structure constant $\alpha$. A notion that all the fundamental couplings of physics are somehow related to one another provides the primary motivation. It seems rational too, given the fact that we have no clear interpretation of their origin or their constant nature. The formulation is inspired by a similar approach of \cite{barrowplb}. We consider a two-scalar extended theory of gravity. One of the scalar fields, $\phi$, is of Brans-Dicke (BD) nature, i.e., geometric and accounts for the variation of the effective gravitational coupling \citep{bransdicke}. The second scalar field is the $e$-field as in a standard BSBM setup. We also introduce a dimensionless BD coupling $\omega_{BD}$. However, keeping in mind the stringent constraints on the value of BD parameter from local astronomical tests, we deliberately take $\omega_{BD}$ to be a function of $\phi$. Such extensions were first considered by \cite{nordt} and since then have received a few revisits in the context of cosmological analysis of modified BD theories \citep{barker, swinger, vdb}. The varying $\alpha$ in this generalized Brans-Dicke setup behaves as an effective matter field and therefore we call the theory a generalized \textit{Brans-Dicke-BSBM (BDBSBM)} setup. The combined action is written as

\begin{equation}
S = \int d^4x\sqrt{-g}\left( R\phi +\frac{16\pi }{c^4}{\cal L}-\omega_{BD}(\phi) \frac{\phi _{,\mu }\phi ^{,\mu }}\phi \right).
\label{act}
\end{equation}

The energy momentum Lagrangian consists of three parts,
\begin{equation}
{\cal L} = {\cal L}_m + {\cal L}_{em}\exp (-2\psi ) + {\cal L}_\psi.  \label{lag}
\end{equation}

The $e$-field contribution to the energy-momentum tensor is written as
\begin{equation}
{\cal L}_\psi =-\frac{\omega(\psi)}{2}\psi _{,\mu}\psi ^{,\mu}.
\end{equation}

As usual, we are interested in a spatially homogeneous cosmology for which the independent equations are written as

\begin{eqnarray}\nonumber
&& 3H^2 = \frac{8\pi }\phi \left( \rho _m(1+|\zeta|\exp (-2\psi ))+\rho _r\exp (-2\psi ) + \rho _\psi \right) \\&&
-3\frac{\dot a^{\ }\dot{\phi}}{a\ \phi }+\frac{\omega _{BD}(\phi)}2\frac{\dot \phi ^2}{\phi ^2}-\frac k{a^2},  \label{fried} \\&&\nonumber
\ddot \phi +3H\dot{\phi} = \frac{8\pi}{3+2\omega_{BD}(\phi)}(\rho_m - 2\rho_\psi) - \frac{\dot{\phi}^{2}}{3+2\omega_{BD}(\phi)}\\&&
\frac{d\omega_{BD}(\phi)}{d\phi},  \label{phi} \\&&
\ddot{\psi} +3H\dot{\psi} + \frac{1}{2\omega(\psi)}\frac{d\omega(\psi)}{d\psi}\dot{\psi}^{2} = -\frac{2 \exp (-2\psi )}{\omega(\psi)}\zeta \rho_{m}.\label{ddotpsi}
\end{eqnarray}

Apart from the BD field and its functions, other terms are quite similar to the case considered in the last section. $\omega(\psi)$ signifies the energy scale of $\psi$ and should not be confused with the Brans-Dicke counterpart $\omega_{BD}(\phi)$. $\rho_\psi$ is the kinetic energy density contribution of $\psi$ and is given by 
\begin{equation}
\rho_\psi = \frac{\omega(\psi)}{2}\dot{\psi}^2.
\end{equation}
$\rho_m$ is proportional to $a^{-3}$. Conservation of the noninteracting radiation density $\rho_r$ leads to the condition
\begin{equation}
\rho_{ir} = \rho_{r}e^{-2\psi} \propto a^{-4}.
\end{equation}

\begin{figure}
\begin{center}
\includegraphics[angle=0, width=0.40\textwidth]{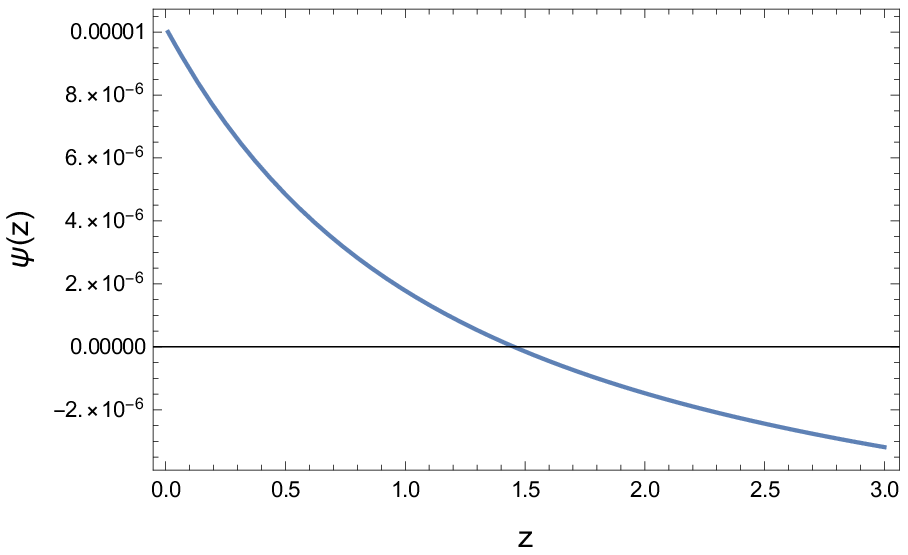}
\includegraphics[angle=0, width=0.40\textwidth]{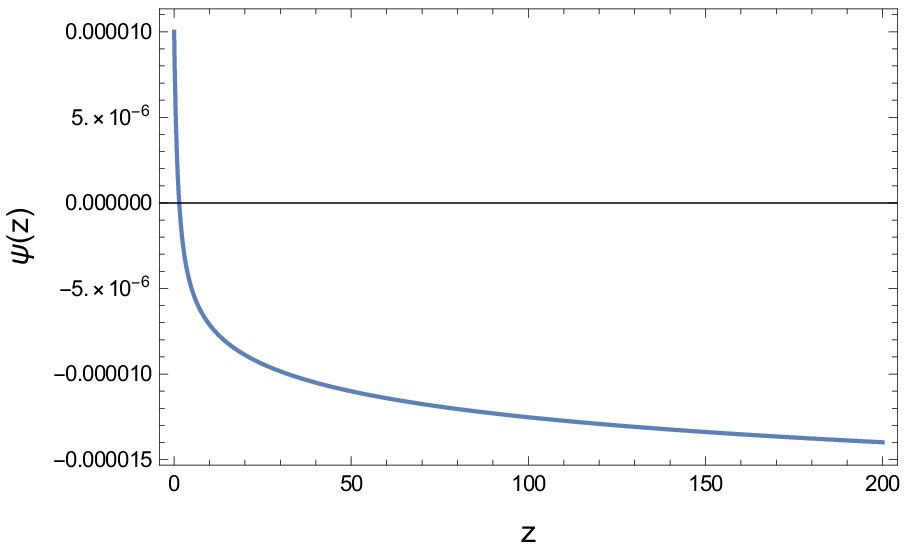}
\caption{Generalized Brans-Dicke-BSBM : Evolution of $\psi$ in the low redshift regime (top panel) and high redshift regime (bottom panel) for the best fit parameter values of $\lambda_{0}$, $\delta$ and $H_{0}$ and $\zeta > 0$.}
\label{scalar_BDBSBM}
\end{center}
\end{figure}

\begin{figure}
\begin{center}
\includegraphics[angle=0, width=0.40\textwidth]{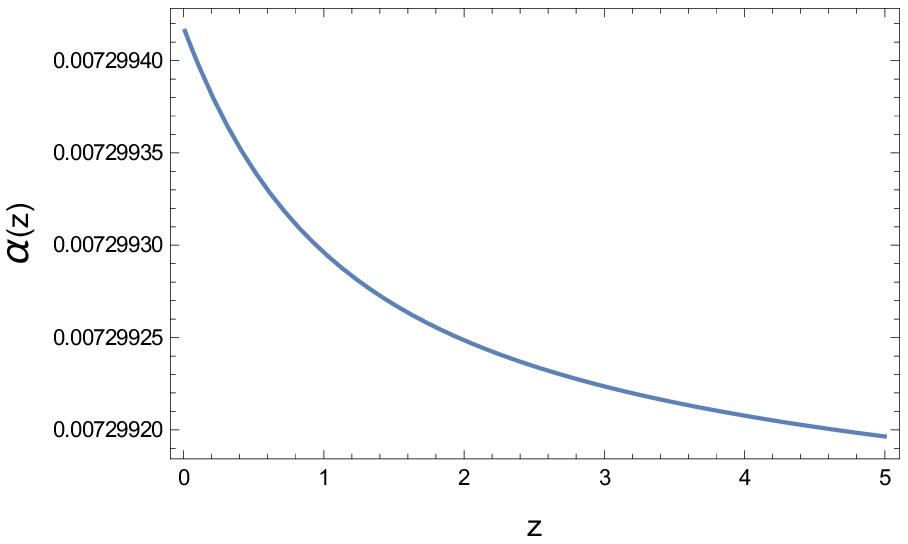}
\includegraphics[angle=0, width=0.40\textwidth]{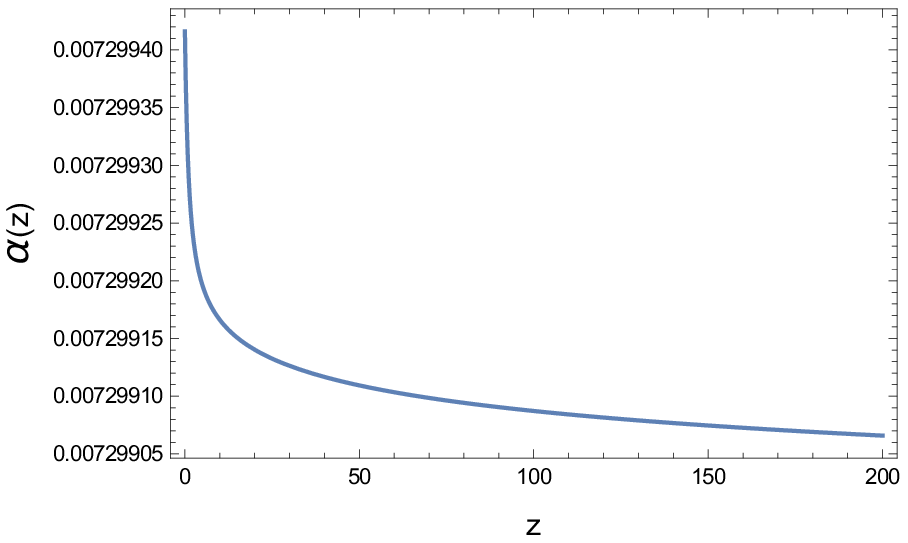}
\caption{Generalized Brans-Dicke-BSBM : The evolution of $\alpha$ with $z$ for the best fit parameter values of $\lambda_{0}$, $\delta$ and $H_{0}$ and $\zeta > 0$. Top Panel shows the evolution for low redshift and the Bottom Panel shows the evolution for high redshift.}
\label{alpha_BDBSBM}
\end{center}
\end{figure}

In a unit $\hbar = c = 1$, the fine structure coupling has a correlation with $\psi$ as

\begin{equation}
\alpha = \alpha_{0} e^{2\psi},  \label{alf}
\end{equation}
$\alpha_0$ being the value of fine structure constant as we know today. We solve the above set of field equations, primarily to derive $\psi$. Since the unknown functions in this case outnumber the independent equations, we choose $\omega(\psi)$ at the outset as
\begin{equation}
\omega(\psi) = \omega_{0}e^{\mu \psi},
\end{equation}

with the constraint that $\omega_{0} > 0$ for all time. We also set the present value of Newtonian coupling $G$ to be unity such that while solving for $\phi$ we have a fixed initial condition ($\phi \propto 1/G$). $\zeta = {\cal L}_{em}/\rho_m$ is the parameter defining non-relativistic matter contribution to $L_{em}$. It demands specific attention for this particular example of extended BSBM as it can establish a correlation between the scalar dynamics and the cold dark matter constituents of our universe. For instance, $\zeta < 0$ corresponds to models where the magnetic energy of the interacting scalar-charged matter system dominates over the electric field energy. Example of Dark matter being dominated by magnetic coupling can be found in superconducting cosmic strings for which $\zeta =-1$. These cases have received some attention in literature and they support a mildly varying $\alpha$ during the dust-dominated era followed by an almost constant $\alpha$ during late times \citep{bsbm, barrowprd}. In comparison, a positive $\zeta$ case is expected contradict with observations of molecular absorption spectroscopy \citep{dvali}. Since there is no clear knowledge of the Dark matter distribution in our universe, we take a diplomatic approach and look into the evolution of $\alpha$, $\Delta\alpha/\alpha$ and the scalar fields for both $\zeta < 0$ and $\zeta > 0$, while always restricting $\zeta$ to be between $-1$ and $+1$. Since we rely on a cosmological reconstruction supported by a set of widely accepted observations, the expansion scale factor is never affected by the $\alpha$ variations. Therefore the choices of $\zeta$ can only modify the $e$-field evolution and do not come at odds with the primary cosmological requirements. \\

\begin{figure}
\begin{center}
\includegraphics[angle=0, width=0.40\textwidth]{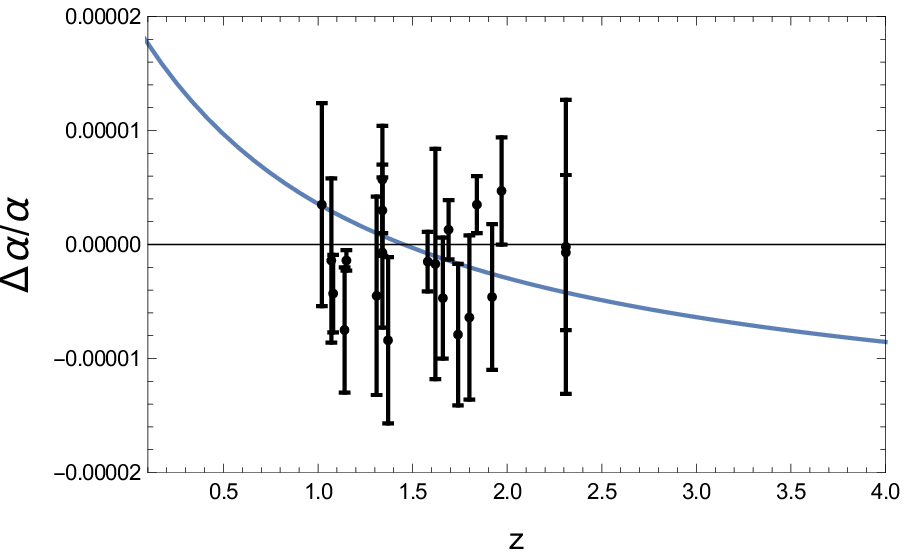}
\includegraphics[angle=0, width=0.40\textwidth]{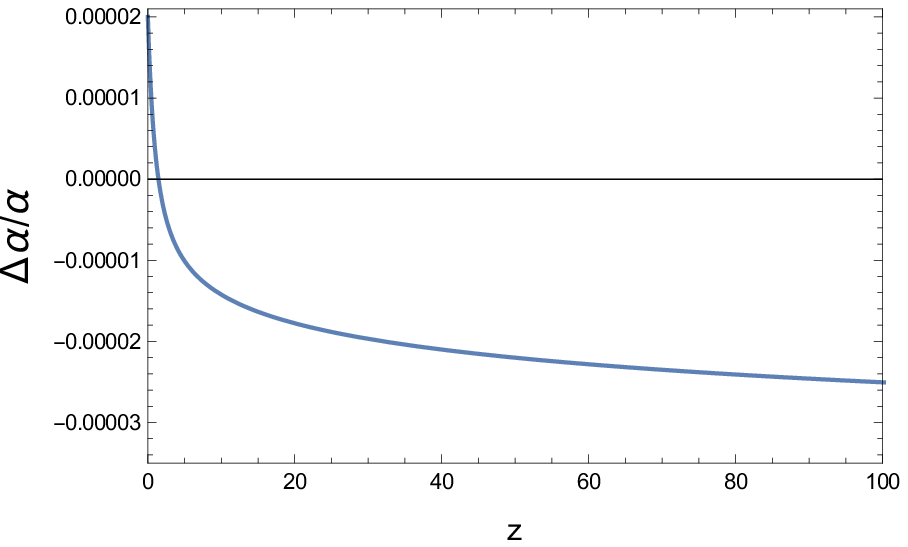}
\caption{Generalized Brans-Dicke-BSBM : The evolution of $\Delta\alpha/\alpha$ with $z$ for the best fit parameter values of $\lambda_{0}$, $\delta$ and $H_{0}$ and $\zeta > 0$. The Top Panel shows a plot for low redshifts alongwith the molecular spectroscopy data. The Bottom panel shows the plot for a larger range of redshifts.}
\label{deltaalpha_BDBSBM}
\end{center}
\end{figure}

\begin{figure}
\begin{center}
\includegraphics[angle=0, width=0.40\textwidth]{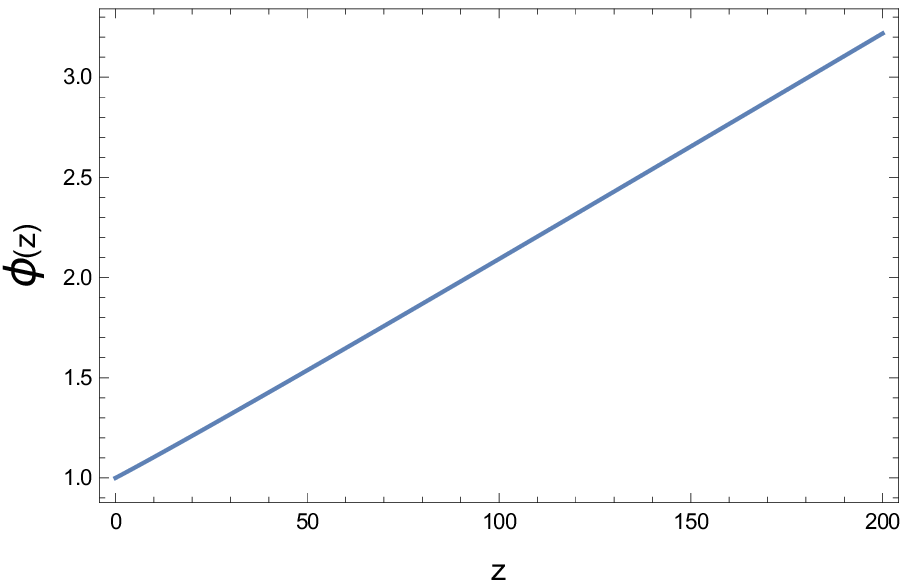}
\includegraphics[angle=0, width=0.40\textwidth]{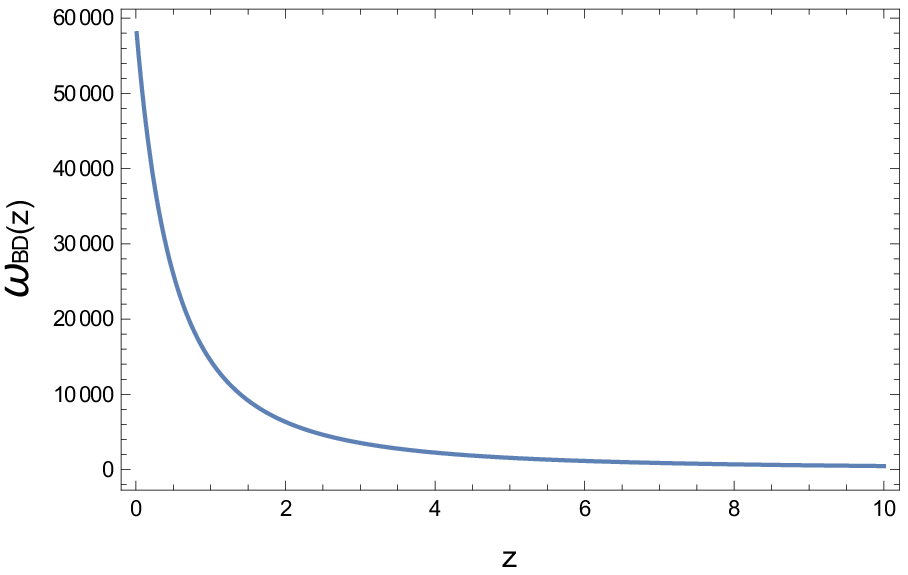}
\caption{Generalized Brans-Dicke-BSBM : The evolution of the Brans-Dicke scalar $\phi$ (Top Panel) and the coupling $\omega_{BD}$ (Bottom Panel) with $z$ for the best fit parameter values of $\lambda_{0}$, $\delta$ and $H_{0}$ and $\zeta > 0$. The initial condition for the numerical solution is taken to be $\dot{\phi} > 0$.}
\label{BD_BDBSBM}
\end{center}
\end{figure}

\begin{figure}
\begin{center}
\includegraphics[angle=0, width=0.40\textwidth]{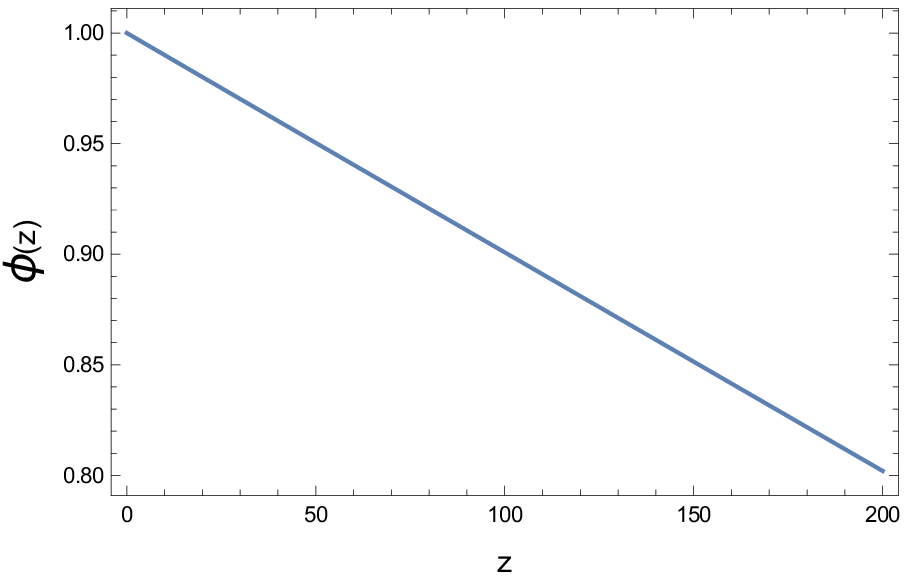}
\includegraphics[angle=0, width=0.40\textwidth]{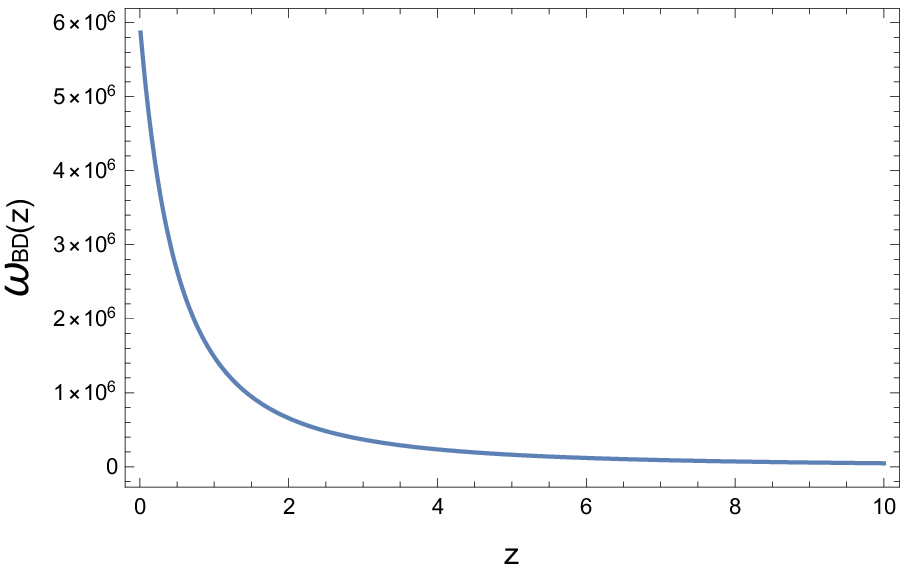}
\caption{Generalized Brans-Dicke-BSBM : The evolution of the Brans-Dicke scalar $\phi$ (Top Panel) and the coupling $\omega_{BD}$ (Bottom Panel) with $z$ for the best fit parameter values of $\lambda_{0}$, $\delta$ and $H_{0}$ and $\zeta > 0$. The initial condition for the numerical solution is taken to be $\dot{\phi} < 0$.}
\label{BD_BDBSBM2}
\end{center}
\end{figure}

The differential Eqs. (\ref{fried}), (\ref{phi}), and (\ref{ddotpsi}) are solved using the reconstructed Hubble,
\begin{equation}
H(z) = H_{0}\left[1 + \lambda_{0} (1+z)^{\delta} \left\lbrace (1+z)^{3} - 1 \right\rbrace\right]^{\frac{1}{2}}.
\end{equation}

We take the best fit parameter values of $\lambda_{0}$, $\delta$ and $H_{0}$ while doing this. Once again, the primary trick is to change the arguments of the equations from cosmic time into redshift and to solve the equations numerically (See Eqs. (\ref{1eq}) and (\ref{2eq}) for reference). First, we solve the set of equations for $\zeta > 0$. The plots in Fig. \ref{scalar_BDBSBM} show the numerical solution for $\psi(z)$. The top panel of the Figure shows the $\psi$ evolution for low $z$. Around the redshift of transition, the scalar field crosses into a positive domain. For most of the earlier epochs, the scalar is mildly evolving and remains in a negative realm. The overall variation is within the range $\sim \left[0.00001,-0.000015\right]$, i.e., quite mild. This can be understood from the bottom panel of Fig. \ref{scalar_BDBSBM} where $\psi$ evolution is shown for larger redshifts. \\

We plot the fine structure constant in Fig. \ref{alpha_BDBSBM} as a function of redshift. Theoretically, $\alpha$ changes exponentially with $\psi$ and should have a mild evolution, confirmed by the Figure. As discussed in the last section, this variation is better understood through the quantity $\Delta \alpha / \alpha = (\alpha_{z}-\alpha_{0}) / \alpha_{0}$. We plot the theoretically estimated $\Delta \alpha / \alpha$ in Fig. \ref{deltaalpha_BDBSBM} and fit with the data points from Table \ref{table2}. It is clear from the top panel that at low redshift, the theory gives a very good fit with the observations of molecular absorption spectra. The bottom panel shows $\Delta \alpha / \alpha$ for larger ranges of redshifts and suggests that the variation eventually becomes negligibly mild. However, we repeat once again that this estimate is primarily based on a low redshift reconstruction and may not be accurate enough for a large redshift dynamics.  \\

\begin{figure}
\begin{center}
\includegraphics[angle=0, width=0.40\textwidth]{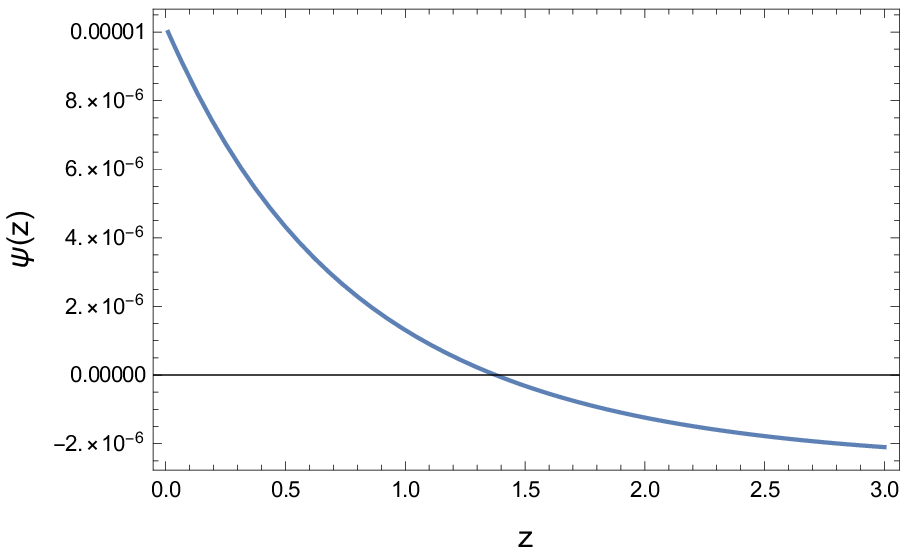}
\includegraphics[angle=0, width=0.40\textwidth]{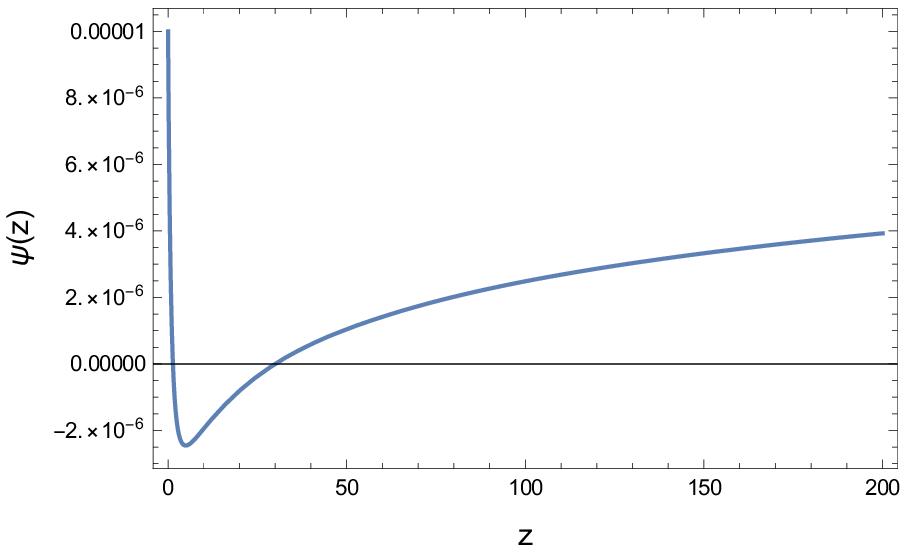}
\caption{Generalized Brans-Dicke-BSBM : Evolution of $\psi$ in the low redshift regime (top panel) and high redshift regime (bottom panel) for the best fit parameter values of $\lambda_{0}$, $\delta$ and $H_{0}$ and $\zeta < 0$.}
\label{scalar_BDBSBM_negativexi}
\end{center}
\end{figure}

\begin{figure}
\begin{center}
\includegraphics[angle=0, width=0.40\textwidth]{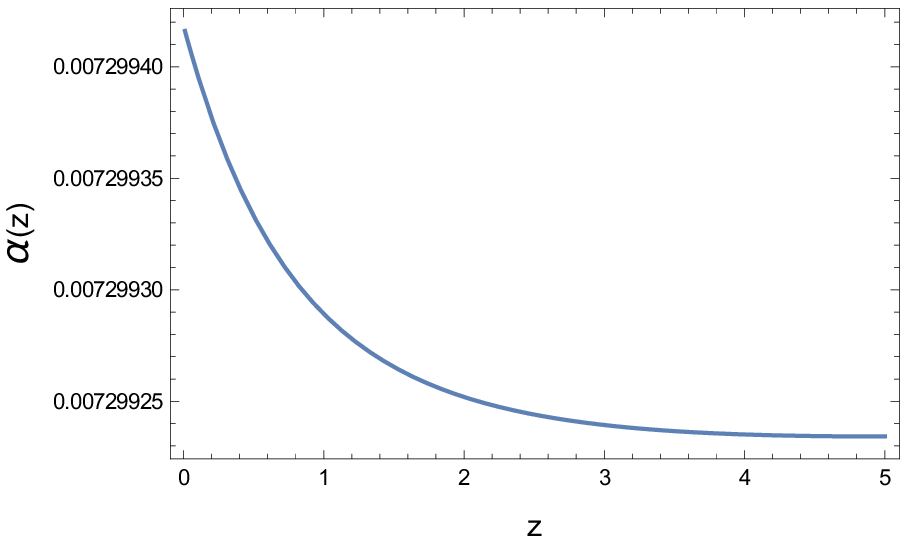}
\includegraphics[angle=0, width=0.40\textwidth]{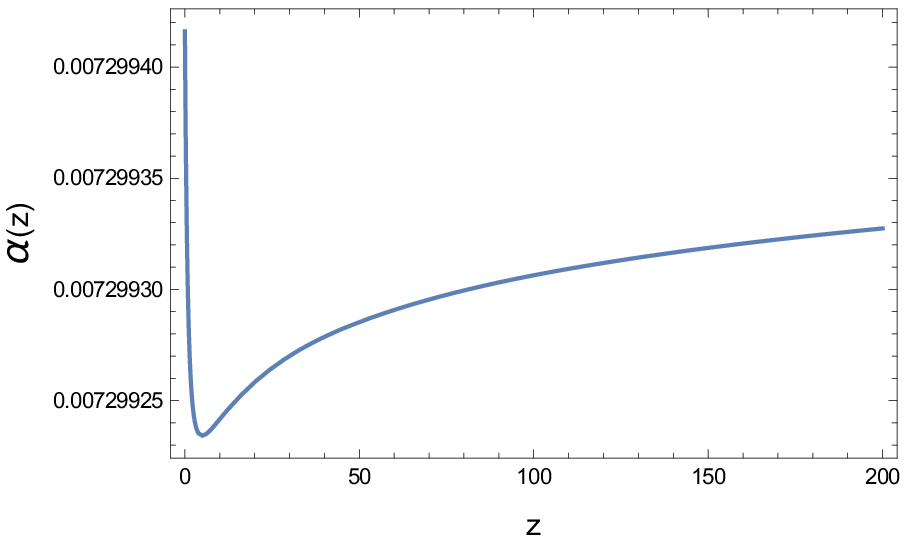}
\caption{Generalized Brans-Dicke-BSBM : Evolution of $\alpha$ in the low redshift regime (top panel) and high redshift regime (bottom panel) for the best fit parameter values of $\lambda_{0}$, $\delta$, $H_{0}$ and $\zeta < 0$.}
\label{alpha_BDBSBM_negativexi}
\end{center}
\end{figure}

The Brans-Dicke part of the theory signifies how the gravitational coupling evolves and we try to give an idea by solving Eq. (\ref{phi}) for $\phi$. The numerical solution depends on the initial conditions $\dot{\phi} > 0$ and $\dot{\phi} < 0$. We plot them in Fig. \ref{BD_BDBSBM} and Fig. \ref{BD_BDBSBM2}. The top panels of both the Figures suggest that $\phi$ evolves almost linearly with $z$. It may increase or decrease with redshift, depending on the choice of $\dot{\phi}$. The bottom panels of the Figures show the evolution of $\omega_{BD}$ as a function of $z$. The $\omega_{BD}(z)$ curves start to grow up around the redshift of smooth transition $z_{t} \sim 1$. It seems that while a large value of the BD parameter may not be essential to drive a deceleration of the universe, it is a necessary requirement to allow the transition into present acceleration. Keeping in mind an almost linear evolution of $\phi$, we intuitively suggest a form of $\omega_{BD}(\phi)$ as
\begin{equation}
\omega_{BD}(\phi) \sim \delta_{1}e^{-(\delta_{2}+\delta_{3}\phi)}.
\end{equation}
The three parameters $\delta_{1}$, $\delta_{2}$ and $\delta_{3}$ should be suitably estimated according to other initial conditions.    \\

We now discuss the structure of the theory that can support a mildly evolving $\Delta\alpha/\alpha$ even for $\zeta < 0$. The numerical solution for $\psi$ is shown in Fig. \ref{scalar_BDBSBM_negativexi}, both for low and larger ranges of redshift. The curve crosses zero twice within the span of evolution. A minima of the scalar field is seen during the matter-dominated deceleration. Prior to this, the scalar field remains in the positive half and sees a mild variation. The $\alpha$ variation is expected to be mild, as shown in Fig. \ref{alpha_BDBSBM_negativexi}. Although the low redshift behavior remains almost similar compared to a positive $\zeta$ case, an interesting departure can be seen in the profile for larger redshifts. \\

\begin{figure}
\begin{center}
\includegraphics[angle=0, width=0.40\textwidth]{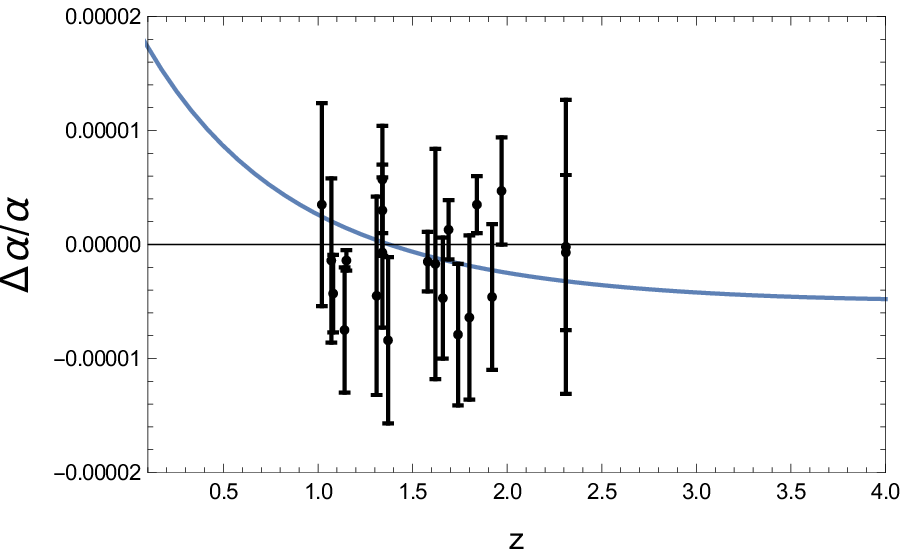}
\includegraphics[angle=0, width=0.40\textwidth]{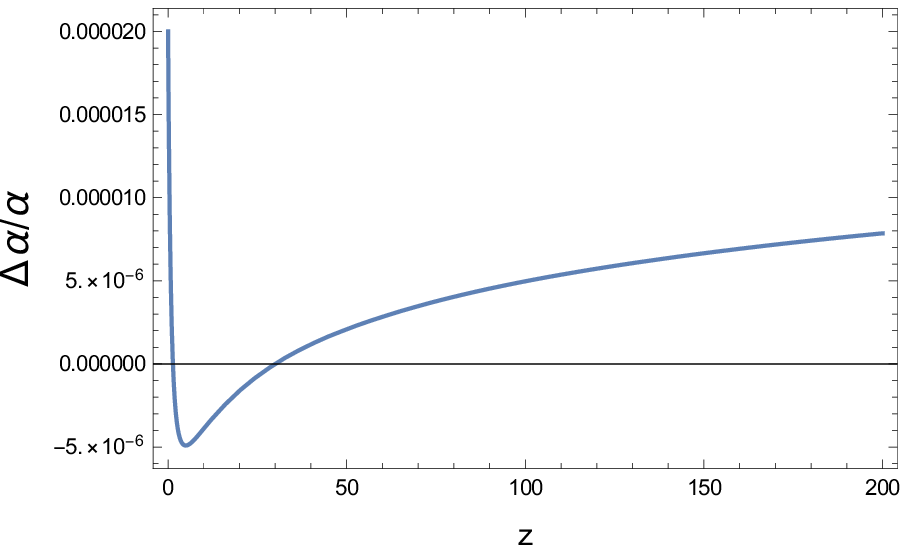}
\caption{Generalized Brans-Dicke-BSBM : The evolution of $\Delta\alpha/\alpha$ with $z$ or the best fit parameter values of $\lambda_{0}$, $\delta$ and $H_{0}$ and $\zeta < 0$. Top Panel shows a plot for low redshifts alongwith molecular spectroscoy data fitted in. The Bottom panel is for larger span of redshifts.}
\label{deltaalpha_BDBSBM_negativexi}
\end{center}
\end{figure}

\begin{figure}
\begin{center}
\includegraphics[angle=0, width=0.40\textwidth]{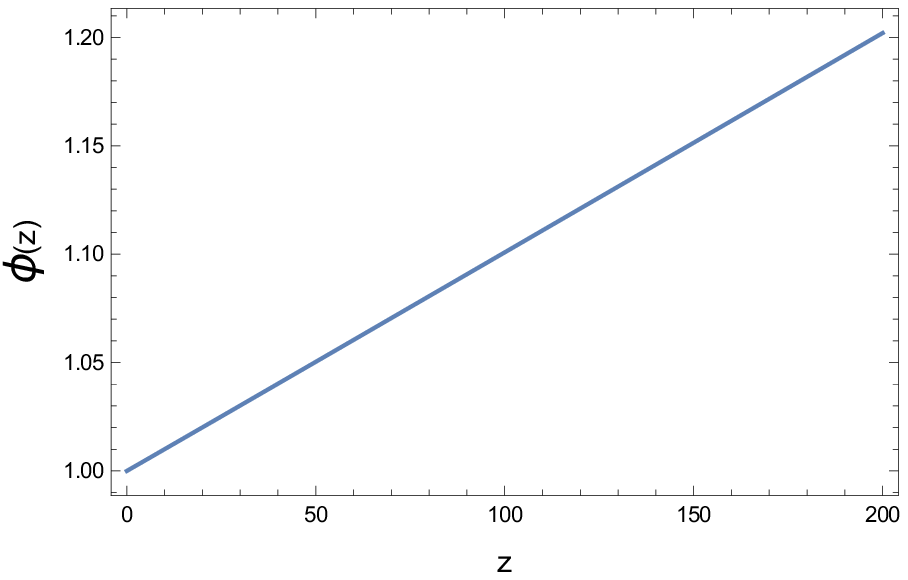}
\includegraphics[angle=0, width=0.40\textwidth]{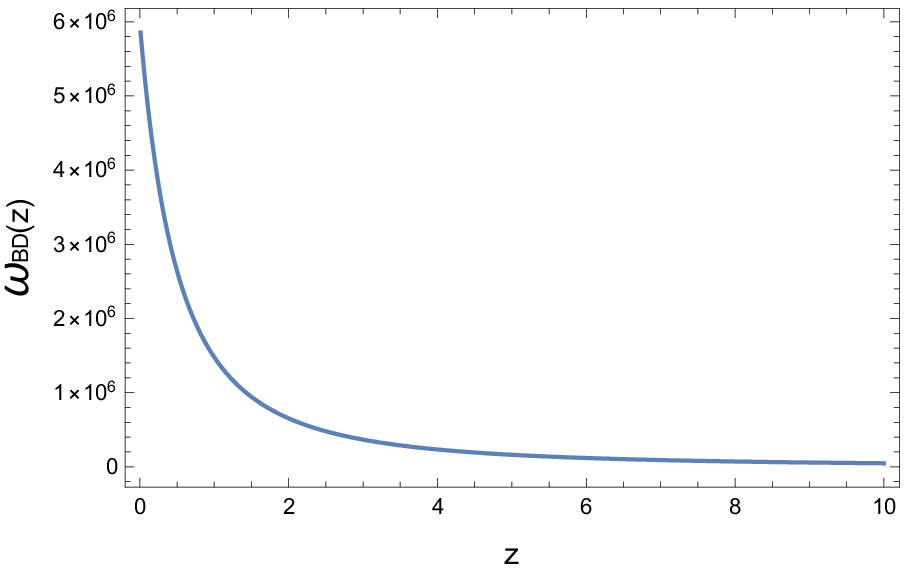}
\caption{Generalized Brans-Dicke-BSBM : The evolution of the Brans-Dicke scalar $\phi$ (Top Panel) and the coupling $\omega_{BD}$ with $z$ (Bottom Panel) for the best fit parameter values of $\lambda_{0}$, $\delta$ and $H_{0}$ and $\zeta < 0$. The initial condition for the numerical solution is taken to be $\dot{\phi} > 0$.}
\label{BD_BDBSBM_negativexi}
\end{center}
\end{figure}

\begin{figure}
\begin{center}
\includegraphics[angle=0, width=0.40\textwidth]{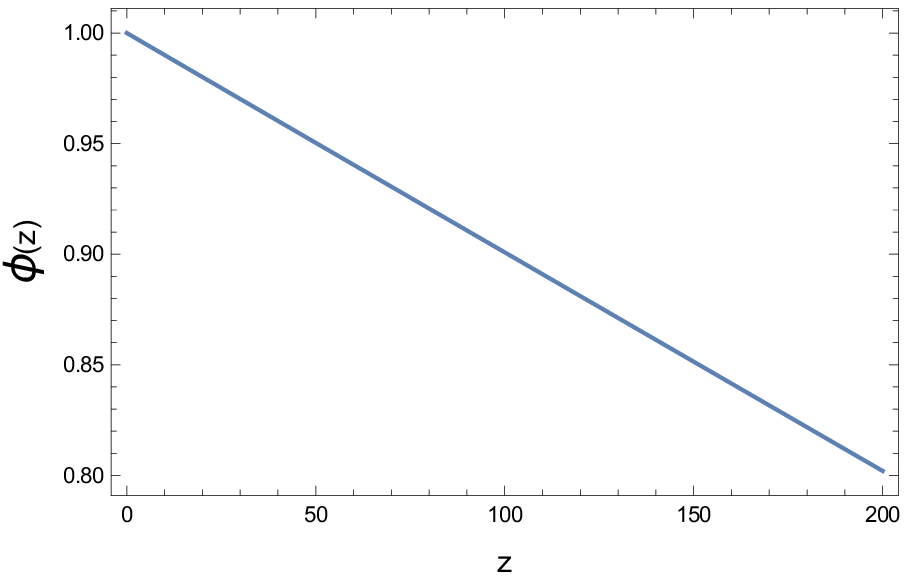}
\includegraphics[angle=0, width=0.40\textwidth]{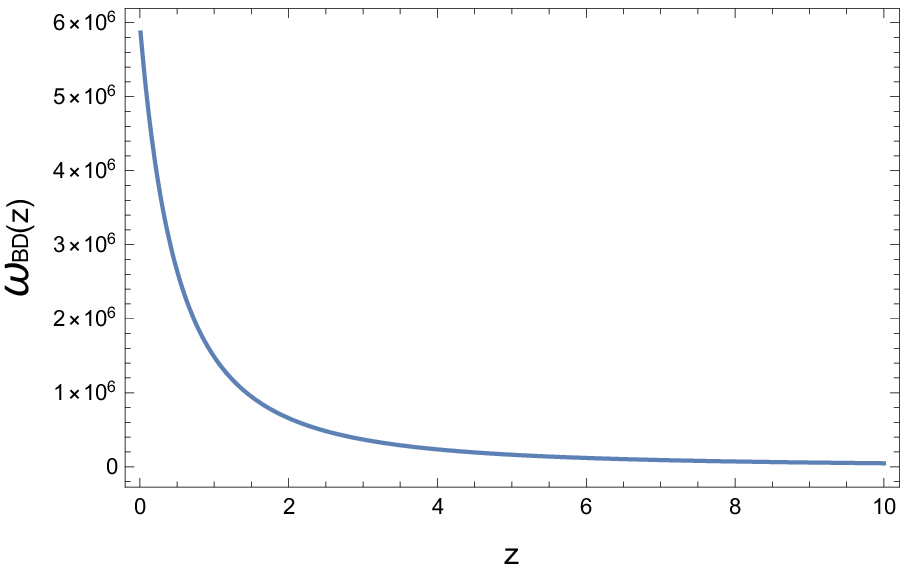}
\caption{Generalized Brans-Dicke-BSBM : The evolution of the Brans-Dicke scalar $\phi$ (Top Panel) and the coupling $\omega_{BD}$ with $z$ (Bottom Panel) for the best fit parameter values of $\lambda_{0}$, $\delta$ and $H_{0}$ and $\zeta < 0$. The initial condition for the numerical solution is taken to be $\dot{\phi} < 0$.}
\label{BD_BDBSBM_negativexi2}
\end{center}
\end{figure}

Nevertheless, the fact that $\psi$ evolves in a different manner for $\zeta < 0$, does not contribute much to the low redshift variation of $\Delta\alpha/\alpha$. This is quite clear from the top panel of Fig. \ref{deltaalpha_BDBSBM_negativexi} where a good fit with the molecular absorption spectroscopic data is once again found. However, for larger ranges of redshift, the evolution is clearly different as compared to a $\zeta > 0$ dynamics, as in the bottom panel of Fig. \ref{deltaalpha_BDBSBM_negativexi}). $\Delta\alpha/\alpha$ evolves in and out of the negative domain within a redshift range of $z \in (5 , 30)$. The formation of this minima coincides with the end of matter domination and it may have a role in setting up the system for the onset of late-time acceleration. We also solve for the BD field $\phi$ and the coupling $\omega_{BD(z)}$ for a $\zeta < 0$ system. The numerical solution is, once again, sensitive on the initial conditions $\dot{\phi} > 0$ and $\dot{\phi} < 0$. These are shown in Fig. \ref{BD_BDBSBM_negativexi} and Fig. \ref{BD_BDBSBM_negativexi2}. The evolution is quite similar to the $\zeta > 0$ cases. Therefore, $\omega_{BD}(\phi) \sim \delta_{1}e^{-(\delta_{2}+\delta_{3}\phi)}$ remains a good possible choice for the generalized BD coupling for all initial conditions under consideration.  \\

\begin{figure}
\begin{center}
\includegraphics[angle=0, width=0.40\textwidth]{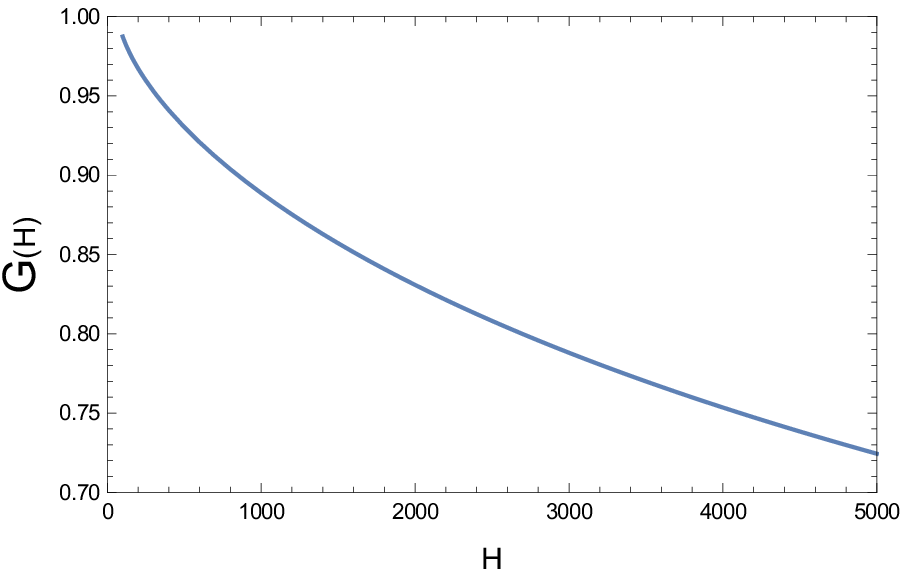}
\includegraphics[angle=0, width=0.40\textwidth]{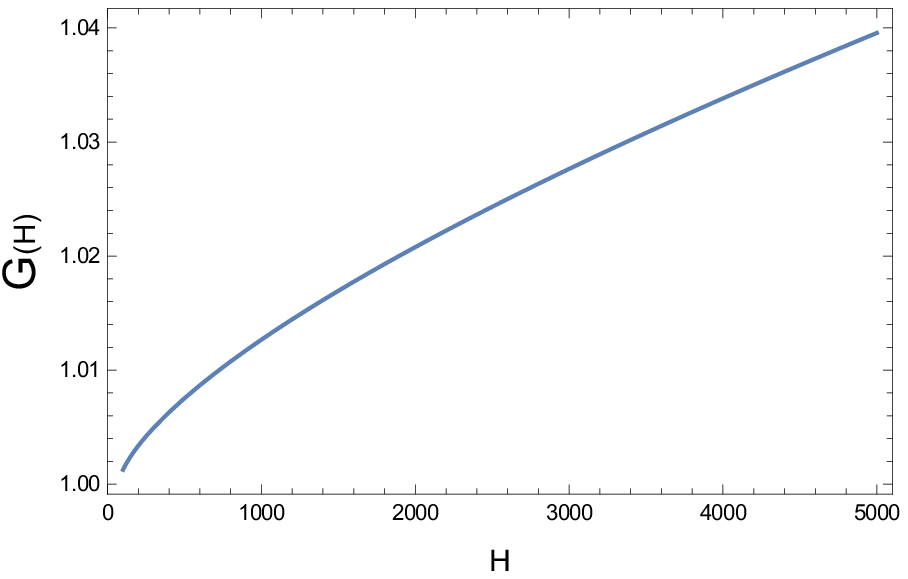}
\caption{Generalized Brans-Dicke-BSBM : Variation of Newtonian Coupling, scaled as $G/G_{0}$ with Hubble for the best fit parameter values of $\lambda_{0}$, $\delta$ and $H_{0}$. $G_{0}$ is the the present value of $G$ (scaled to $1$ in natural units). Top Panel : Plot for $\dot{\phi} > 0$. Bottom Panel : Plot for $\dot{\phi} < 0$.}
\label{GvsH}
\end{center}
\end{figure}

Since these models are special cases of generalized Brans-Dicke theory, the evolution of $G_{eff} \propto 1/\phi$ is an important factor that determines the nature of the theory. Generally, no pre-assigned expectation regarding this evolution can be found in literature, except, the novel bound given by \cite{weinberg}
\begin{equation}
\frac{\dot{G}}{G} = -\frac{\dot{\phi}}{\phi} = +\frac{k}{H}.
\end{equation}
$k \leq 1$. For a scalar-tensor theory this evolution is solution-dependent and we have two different $\phi$ profiles, as in Fig. \ref{BD_BDBSBM} and Fig. \ref{BD_BDBSBM2}. We plot $G_{eff}$ against Hubble, in Fig. \ref{GvsH}. The top panel shows the evolution for the initial condition $\dot{\phi} > 0$. In this case, $G_{eff}$ decays smoothly with Hubble, indicating that gravitational interaction was weaker in the past, when Hubble or the natural scale of energy was higher. The maximum allowed value of $G_{eff}$ is equal to the present value $1$. This particular version of the theory can be called an \textit{`asymptotically free theory'}. However, if $G_{eff}$ decays with Hubble, i.e., increases with cosmic expansion, we can not rule out a probable conflict with the $H_0$ tension on some scale \citep{banerjee, heisenberg, lee}. The second scenario with $\dot{\phi} < 0$ might help us avoid this issue. In this case, due to a monotonically decreasing profile of $\phi$, $G_{eff}$ increases with Hubble. This indicates a decay of the gravitational coupling with cosmic time until the minimum allowed value $1$ is reached. This avoids a further complicated $H_{0}$ tension at the expense of giving away the asymptotically free nature, which may compromise some of the standard model phenomenology \citep{sola3}. We curiously note that, (i) the present epoch enjoys an extremum (either the maxima or the minima) of $G_{eff}$ and (ii) $\zeta$ leaves no contribution in these evolutions. \\

The models given in this section do have some similarities in structure with theories where a \textit{`dynamical vacuum'} can drive a variation in proton-to-electron mass ratio \citep{cruzperez, sola1}. Both of these models support a similar pattern in the mild variation of natural couplings. The connection is potentially intriguing as the dynamical vacuum models are closely related to a cosmic variation of Higgs vacuum expectation value \citep{sola2, sola3, chakrabarti2}. This, again indicates a direct feedback of the theory on the Quark masses and overall, on particle physics phenomenology. We also mention here that a variation of $\frac{\Delta \alpha}{\alpha}$ already allows a $\mu$-variation through 
\begin{equation}
\frac{\Delta\mu}{\mu} \sim \frac{\Delta\Lambda_{QCD}}{\Lambda_{QCD}} - \frac{\Delta{\nu}}{\nu} 
\sim R\frac{\Delta \alpha}{\alpha}. \label{defR}
\end{equation}
Here $R < 0$ is a model-dependent parameter and can be estimated from high energy experiments of unified theories \citep{avelino}. Overall, these analogies motivate a requirement to merge extended theories of gravity with phenomenologies of particle physics. \\

\begin{figure}
\begin{center}
\includegraphics[angle=0, width=0.40\textwidth]{delta_alpha_z_negativexi.eps}
\includegraphics[angle=0, width=0.40\textwidth]{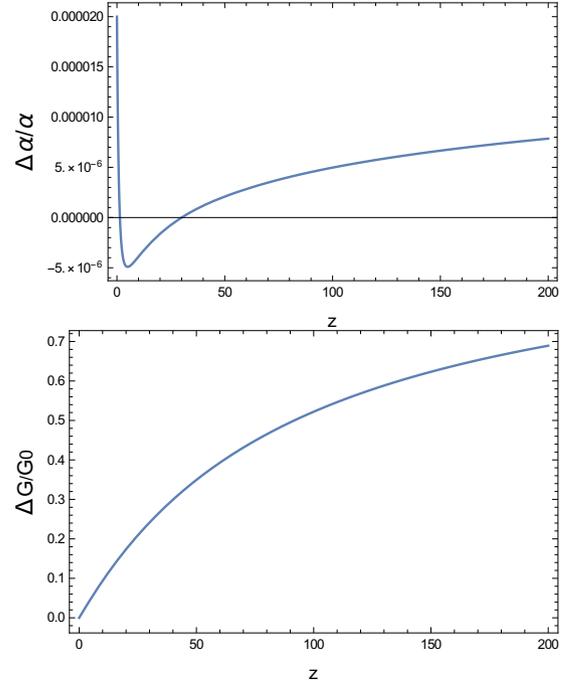}
\caption{Generalized Brans-Dicke-BSBM : A comparison between variations of the fine structure coupling, $\Delta\alpha/\alpha$, and the gravitational coupling $\Delta G/G_{0}$. $G_{0}$ is the present value of $G$, scaled to $1$ in natural units. The plots are for the best fit parameter values of $\lambda_{0}$, $\delta$ and $H_{0}$.}
\label{deltaalphaanddeltaG}
\end{center}
\end{figure}

\begin{figure}
\begin{center}
\includegraphics[angle=0, width=0.40\textwidth]{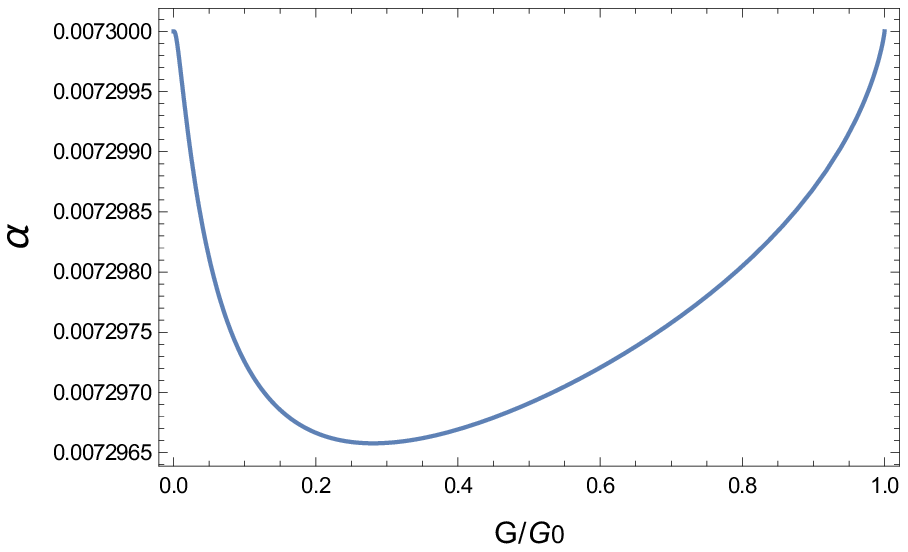}
\includegraphics[angle=0, width=0.40\textwidth]{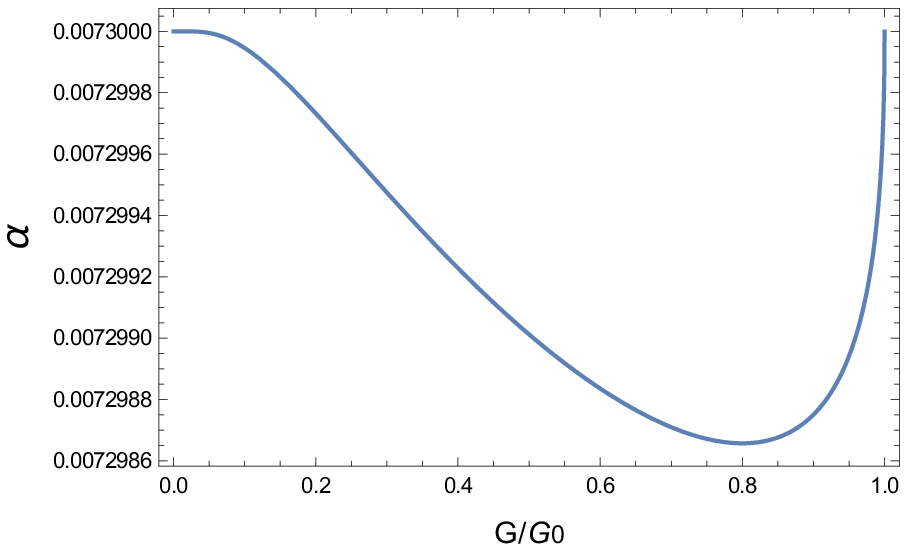}
\caption{Generalized Brans-Dicke-BSBM : The evolution of $\alpha$ vs $G/G_{0}$ for the best fit parameter values of $\lambda_{0}$, $\delta$ and $H_{0}$. $G$ is scaled by $G_{0}$, the present value of $G$ (scaled to $1$ in natural units). The top panel is for $\zeta > 0$ and the bottom panel is for $\zeta < 0$. The initial condition $\dot{\phi} > 0$ is enforced.}
\label{alphavsG}
\end{center}
\end{figure}

Before concluding the section, we briefly talk about an interesting possibility. It is quite reasonable to imagine that more than one fundamental couplings of physics should vary simultaneously and constraining any of them can, inadvertently, constrain the others. The generalized Brans-Dicke-BSBM is an example of the same, where a dimensionless $\alpha$ and a dimensionful $G$ vary alongside each other. The comparative scale of these variations can be realized from Fig. \ref{deltaalphaanddeltaG}, where, $\Delta\alpha/\alpha$ and $\Delta G/G_{0}$ are drawn on the top and the bottom panel respectively, as functions of redshift. Note that, the gravitational coupling $G$ is scaled by $G_{0}$ and written as a dimensionless quantity. $G_{0}$ is the present value of $G$ ($1$ in natural units) which contains all the dimensional informations. This scaling is indeed necessary, in reference to the notion that a concept of \textit{varying natural constant} is rational if and only if it is dimensionless \citep{dirac1, dirac2, duff1, duff2}. Similar variations of more than one fundamental couplings have been addressed to some extent in different scenarios, for instance, in primordial nucleosynthesis \citep{campbell, coc}. Some constraints on a coupled variation can also be determined using Optical atomic clocks \citep{luo, ferreira4}. However, the possibility that one coupling may simply vary as a function of another due to an independent background mechanism, has never been considered. In a recent research on the concurrent variation of $G$, the speed of light $c$, the Planck constant $h$, and the Boltzmann constant $k_B$, the need for an alternative cosmolgical setup was discussed \citep{gupta}. If such a setup exists, it should generate from a fundamental, unified theory. At this moment no such concrete theory is known and we shall have to be content with intuitions, based on whatever evidences we can find. We compare the numerical solutions of fine structure constant $\alpha$ and gravitational coupling $G$ and plot one of them as a function of the other. In Fig. \ref{alphavsG} we plot $\alpha$ vs $G/G_{0}$ for the two signs of $\zeta$. The condition $\dot{\phi} > 0$ is assumed which means that a $G_{eff}$ decaying with Hubble is chosen for this plot. Therefore the allowed range of $G_{eff}$ is $(0, 1]$. We take particular note of the formation of a minima in the evolution of $\alpha$. An early universe is signified by higher energy scales, or higher Hubble, implying lower values of $G$ (See Fig. \ref{GvsH}). $G$ increases with Hubble to the present value, which is scaled to $1$. The two end-points of the plot give two maximum allowed values of $\alpha$ during the cosmic expansion. We speculate that the universe evolves maintaining a correlation with this pattern. It starts evolving with rapid early acceleration where one maxima of $\alpha$ is found. The evolution gradually leads the universe towards the minima of $\alpha$ which coincides with an extended epoch of deceleration. Finally, the universe moves back into the recent acceleration with $\alpha$ approaching the second maxima. The two panels in the Figure are for different signs of $\zeta$.  \\

\begin{figure}
\begin{center}
\includegraphics[angle=0, width=0.40\textwidth]{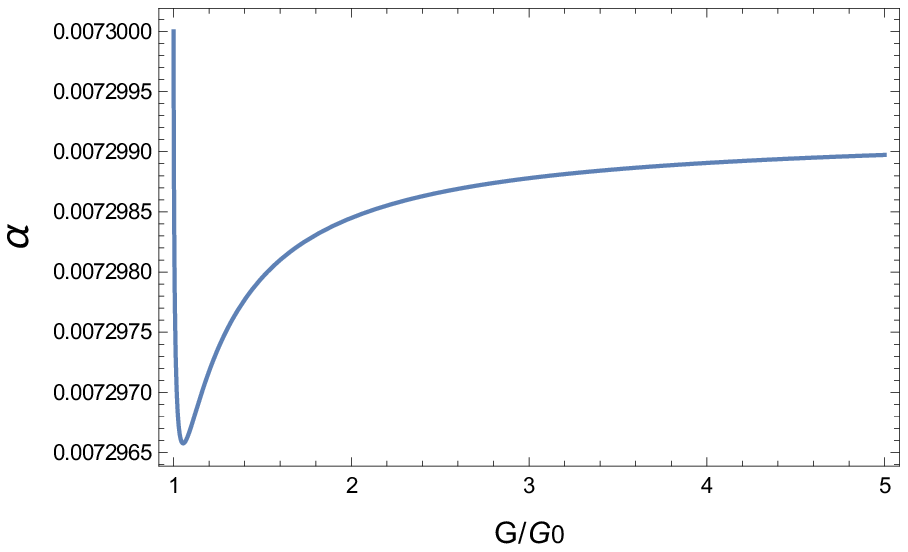}
\includegraphics[angle=0, width=0.40\textwidth]{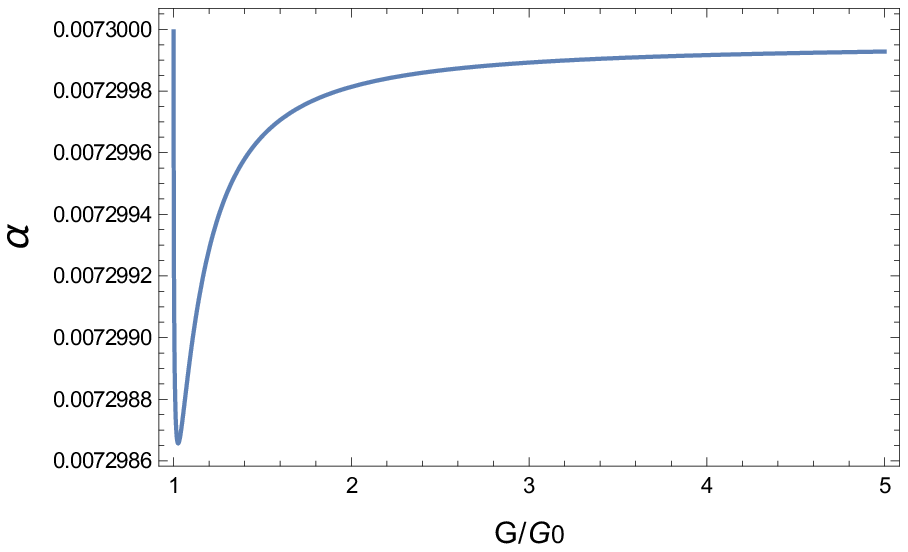}
\caption{Generalized Brans-Dicke-BSBM : The evolution of $\alpha$ vs $G/G_{0}$ for the best fit parameter values of $\lambda_{0}$, $\delta$ and $H_{0}$. $G$ is scaled by $G_{0}$, the present value of $G$ (scaled to $1$ in natural units). The top panel is for $\zeta > 0$ and the bottom panel is for $\zeta < 0$. The initial condition $\dot{\phi} < 0$ is enforced.}
\label{alphavsG2}
\end{center}
\end{figure}

In Fig. \ref{alphavsG2} we plot $\alpha$ vs $G/G_{0}$ for the two signs of $\zeta$, taking $\dot{\phi} < 0$. This case is for the version of the theory where gravitational interaction was stronger in past and decays with cosmic time. $G_{eff}$ can in principle vary in the range $[1, \infty)$ in this case. Quite similar to the previous version of the theory, a clear formation of minima is noted. $\alpha$ evolves negligibly in the cosmological past (i.e., for $G_{eff} > 1$) and rolls along the variation of $G_{eff}$. The two end-points of the graph, i.e., two maximum allowed values of $\alpha$ can mark the epochs of cosmic acceleration, while the minima signals an epoch of deceleration. The overall scale of the variation is quite mild. However, around the present epoch where $\alpha$ starts growing in a rather dominant manner, which is a bit contrary to physical expectations. \medskip

Depending on a combination of calculative guess and basic fitting we give a rough functional form that describes how $\alpha$ might have evolved with $G$,
\begin{eqnarray}\nonumber
&&\alpha = \alpha_{0} \Bigg[1-\gamma_{0}\Bigg\lbrace (-1)^{\varepsilon}\gamma_{1} \Bigg(\frac{G_{0}}{G}-1\Bigg)\Bigg\rbrace^{\gamma_{2}}\\&&
e^{{-\gamma_{3}\Big\lbrace (-1)^{\varepsilon}\gamma_{1} \Big(\frac{G_{0}}{G}-1\Big)\Big\rbrace}^{\gamma_{4}}}\Bigg].
\end{eqnarray}

$\alpha_{0}$ is the present value of the fine structure coupling, which is close to $\simeq 0.0073$. The ratio $\frac{G_{0}}{G}$ is dimensionless, where $G_0$ is the present value of the coupling.
\begin{itemize}
\item {$\varepsilon = \pm 1$. This parameter indicates the nature of the theory under consideration. For $\varepsilon = + 1$, $G_{eff}$ decays with Hubble, i.e., grows with cosmic expansion. For $\varepsilon = - 1$, $G_{eff}$ grows with Hubble, i.e., decays with cosmic expansion.}
\item {For $\zeta > 0$, we estimate that (i) $\gamma_{0} \simeq 0.0001$, (ii) $\gamma_{1} \simeq 100$, (iii) $\gamma_{2} \simeq 1$, (iv) $\gamma_{3} \simeq 1$ and (v) $\gamma_{4} \sim 0.25$.}
\item {For $\zeta < 0$, we estimate that (i) $\gamma_{0} \simeq 0.0001$, (ii) $\gamma_{1} \simeq 100$, (iii) $\gamma_{2} \simeq 0.5$, (iv) $\gamma_{3} \simeq 0.2$ and (v) $\gamma_{4} \sim 0.5$.}
\end{itemize}

Since we are talking about the evolution of the universe in terms of look-back time only, an extended question in this regard would be to ask whether the universe will continue to follow this pattern in the future as well. This is a bit difficult to answer based solely on this analysis. We will need a dynamical system analysis of the differential equations governing the system, in order to identify the attractor fixed points. Moreover, the nature and distribution of the cold dark matter does not interfere much in this issue. Nevertheless, it can affect the slope of the curves which means it can affect the onset and span of different epochs as well as the rate of expansion, through the parameter $\zeta$. The manner in which the universe evolves is never affected. 

\section{Conclusion}
Fundamental couplings are an essence of the most rudimentary postulates of physics. However, their origin remains an enigma. Many decades have passed since it was first conjectured that these couplings can characterize different states of the evolving universe and should be treated as varying entities. Today the research in this arena is quite rich, combining non-trivial theoretical formulations with advanced experimental/observational data. We contribute to this genre by considering a generalized theory of varying fine structure constant $\alpha$. The theory is inspired from a generally covariant formalism originally given by \cite{bekenstein} where $\alpha$ varies due to a real scalar field interacting with charged matter. The cosmological extension of Bekenstein's theory, developed by \cite{bsbm}, is popular as a BSBM setup. We work with two examples where a standard BSBM is further generalized. The first generalization allows a field dependent-kinetic term. The second one is a unified theory of two simultaneously varying constants, $\alpha$ and the gravitational coupling $G$. This is done by including a Brans-Dicke type geometric scalar in the Lagrangian. Kinetic terms for both of the scalars are functions of the respective fields in this example. Our aim is to explore the required structure of these extended theories such that a consistent cosmological evolution can be realized alongwith the mild variation of $\alpha$. \\

We utilize a simple methodology for this exercise, based on a parametrization of the present matter density of the universe, written as $Om(z)$. It gives us the Hubble function in closed form, simple enough to compare the theories with reasonable sets of observational data. The observations are of the present epoch and therefore this reconstruction leads to a consistent late-time cosmology, describing the smooth deceleration-to-acceleration transition very nicely. Moreover, we do not need to assume any form of a Dark energy EOS at the outset. Through a statistical analysis of the cosmological data and the comparison with theoretical calculations we establish the validity of the model. We also bring in specific measurements of $\alpha$ variations, reported as ${\Delta\alpha}/{\alpha}$ vs redshift in the spectroscopic analysis of molecular absorption lines observed at Keck and VLT telescopes. A careful comparison of these observations with the theoretically derived $\Delta\alpha/\alpha$ provides us comprehensible constraints on the allowed variation of $\alpha$ and on the extended theories as well. \\

\begin{figure}
\begin{center}
\includegraphics[angle=0, width=0.40\textwidth]{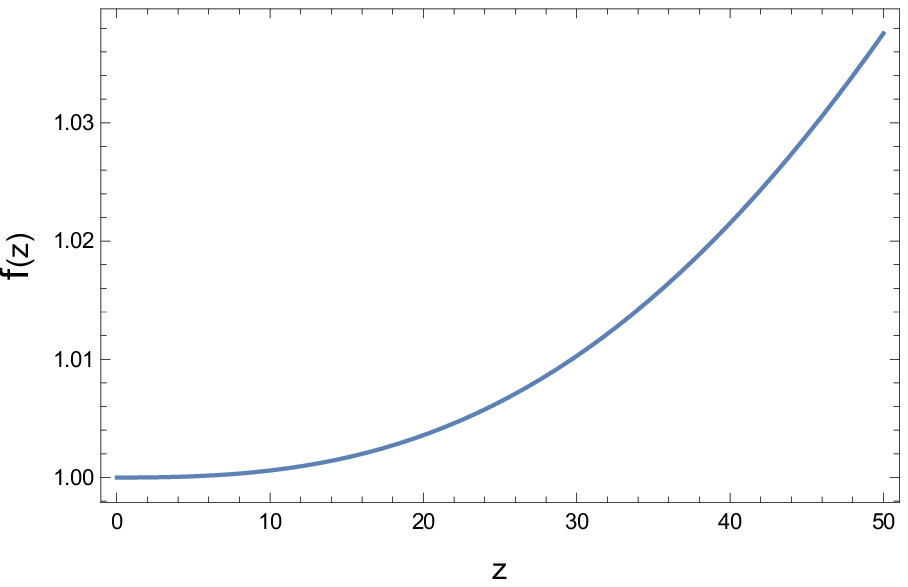}
\includegraphics[angle=0, width=0.40\textwidth]{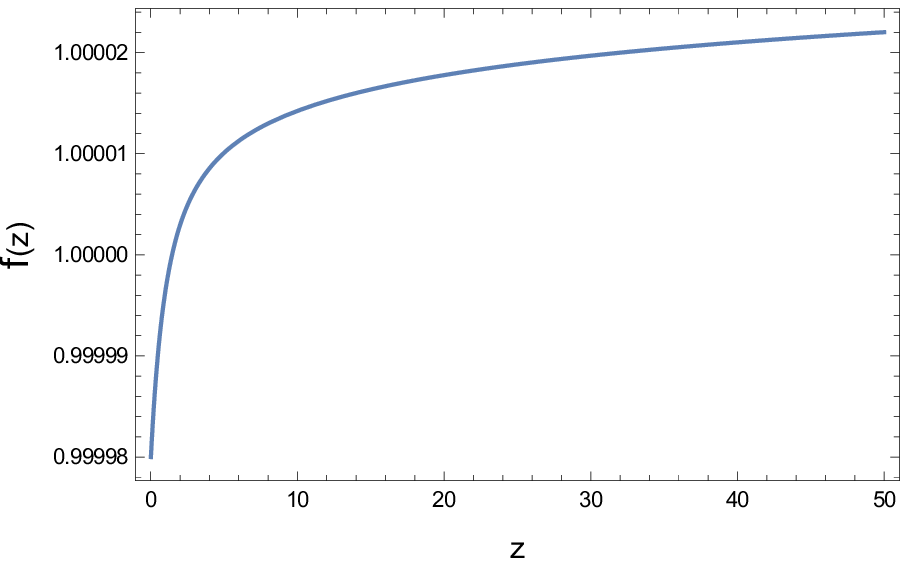}
\caption{The evolution of scalar-matter interaction $f(\psi) \sim e^{-2\psi}$ with $z$ for the best fit parameter values of $\lambda_{0}$, $\delta$ and $H_{0}$. Top Panel : Interaction for model I, generalized BSBM. Bottom Panel : Interaction for model II, generalized BDBSBM.}
\label{interactionfig}
\end{center}
\end{figure}

The scalar field responsible for $\alpha$ variation is of chameleon nature. It interacts with matter in the Lagrangian and acquires a density-dependent mass term. This is realized from the Lagrangian Eq. (\ref{standbsbn}) and the scalar Eq. (\ref{boxeq}). As a result, the scalar can decouple around massive objects due to its own interaction with matter and avoid detection. In extension, this might contribute to the emerging questions involving a violation of Equivalence Principle Constraints. With recent observations disfavoring the case of a standard chameleon field driving the cosmic expansion, the requirement for generalized formalisms such as these, increases manyfold. The models in this manuscript, in a sense, provides two different examples, (i) by generalizing the kinetic part of a chameleonic field and (ii) by including a second field of geometric origin. We note in passing that the scalar-matter interaction profiles ($f(\psi) \sim e^{-2\psi}$) for these two models are found to be quite different as a function of redshift, although the overall variation of there profile is mild (See Fig. \ref{interactionfig} for reference). \\

Both of the models give a mild variation of $\alpha$ consistent with observations. The first model is a natural generalization with the kinetic part evolving roughly as an exponential of the $e$-field (See Fig. \ref{omegapsi} for reference), $\omega(\psi) \sim e^{\beta}\psi$. The numerical solutions of the field equations suggest that $\beta$ should be positive, i.e., $\omega(\psi)$ should grow with $\psi$ for a consistent late-time cosmology. This is well consistent with the results of \cite{barrowlip}, although here the form is derived from cosmological requirements and not assumed at the outset. For the second model, we are forced to choose a similar form of $\omega(\psi)$ to be able to solve the non-linear equations. The rate of variation of $\Delta\alpha/\alpha$ is a bit different for this model, resulting in a nicer fit with observations. Additionally, this is a special example of generalized Brans-Dicke theory with the varying $\alpha$ being fitted in as an effective matter field. The kinetic coupling $\omega_{BD}$ is a function of the Brans Dicke scalar $\phi$ and its evolution is also determined by solving the field equations. We have also guessed a most likely functional form of $\omega_{BD}(\phi) \sim \delta_{1}e^{-(\delta_{2} + \delta_{3}\phi)}$ by fitting in with the numerical solution. For a standard Brans-Dicke theory, $\omega_{BD}$ needs to be quite a large number to provide a viable cosmological solution. With the present generalization, we say that the function $\omega_{BD}(\phi)$ needs to behave differently for different epochs, as per requirements. It dominates since an era of late-time acceleration sets off but is subdued throughout the deceleration. This model also shows how the cold dark matter distribution in the present universe can contribute to the nature of $\alpha$ evolution, through the parameter $\zeta = {\cal L}_{em}/\rho_m$. The signature of $\zeta$ determines the ratio of magnetic energy to the electric field energy of the system. We show that while the low redshift behavior of $\psi$, $\alpha$ and $\Delta\alpha/\alpha$ do not show much of a difference between $\zeta < 0$ and $\zeta > 0$, for higher redshifts the evolutions are characteristically distinct. \\

The generalized Brans-Dicke-BSBM model assembles two separate natures of a scalar-tensor theory into one fold. These models allow a varying $G_{eff} \propto 1/\phi$, where $\phi$ is a geometric scalar field. Depending on the $\phi$ profile, the theory can illustrate two different characters. For a monotonically increasing $\phi$, $G_{eff}$ decays smoothly with Hubble, indicating that gravitational interaction was weaker in the past. On the other hand, for a monotonically decreasing $\phi$, $G_{eff}$ increases with Hubble, i.e., decays with cosmic time. It is, however, noted that the latter of these two cases may further complicate the already non-trivial issue of $H_0$ tension. A weakening $G_{eff}$ during the present cosmic acceleration can avoid this complication, but only at the expense of giving away the asymptotically free nature of the theory. We also find that the present epoch enjoys an extremum of $G_{eff}$ irrespective of the nature of the theory, independent of the initial conditions. \\

We conclude with a hope that the generalized Brans-Dicke-BSBN formalism in particular, can be thought of as the special case of a more fundamental theory. This theory, if formulated in a better way, should be able to describe the variations of all the fundamental couplings within one mathematical construct. This postulation finds motivation from a comparative analysis of the variations of two natural couplings, the fine structure constant and the gravitational coupling. The pattern of their mutual variation may as well shed some light on the phases of cosmic expansion. (i) The universe starts evolving from an early phase of acceleration where there is an extremum of $\alpha$ and $G$. (ii) The second phase sees $\alpha$ rolling down as a function of $G$, towards a minima and this coincides with an epoch of extended deceleration. (iii) Finally, $\alpha$ rolls back up to a second extremum with the universe entering the phase of recent acceleration. $G$ is either a monotonically increasing or a monotonically decreasing function of Hubble, illustrating two entirely different kind of theories within the same framework. We put forward a general prediction that any epoch of cosmic acceleration should exhibit an extremum of any natural coupling.   Although the cold dark matter distribution of the universe can affect the onset and span of different epochs, it has negligible effect on the $\alpha$ vs $G$ pattern. This pattern may not be completely comprehensive, however, the analogies do motivate the requirement for a unified field theory; perhaps through a consistent assembly of extended theories of gravity, phenomenologies of particle physics and a better implementation of the Large Numbers hypothesis. \\

{\bf Data Availability Statement} This manuscript has no associated data or the data will not be deposited.  \\

\section{Acknowledgement}
The author thanks Prof. Michael Duff (Emeritus Professor, Blackett Laboratory, Imperial College London), Prof. Eoin Colgain (Sogang University), Prof. Shahin Sheikh-Jabbari (School of Physics, IPM, Tehran) and Prof. Rajendra Gupta (Adjunct Professor, Department of Physics, University of Ottawa) for their judicious comments on the manuscript. The author also thanks Prof. Koushik Dutta (IISER Kolkata) and Prof. Amitabha Lahiri (SNBNCBS Kolkata) for their thoughtful suggestions.

\bibliographystyle{unsrt}

\begin{thebibliography}{}

\bibitem[\protect\citeauthoryear{Adelberger, Heckel and Nelson}{2003}]{adel}
Adelberger E. G., Heckel B. R. and Nelson A. E., 2003, Ann. Rev. Nucl. Part. Sci. 53, 77.

\bibitem[\protect\citeauthoryear{Agafonova, Molaro, Levshakov and Hou}{2011}]{agafonova}
Agafonova I. I., Molaro P., Levshakov S. A. and Hou J. L., 2011, A and A, 529, A28.

\bibitem[\protect\citeauthoryear{Alam, Sahni, Saini and Starobinsky}{2003}]{alam}
Alam U., Sahni V., Saini T. D. and Starobinsky A. A., 2003, Mon. Not. R. Astron. Soc. 344, 1057.

\bibitem[\protect\citeauthoryear{Albrecht and Magueijo}{1999}]{albrecht}
Albrecht A. and Magueijo J., 1999, Phys. Rev. D. 59 , 043516.

\bibitem[\protect\citeauthoryear{Antoniadis}{1999}]{antoniadis}
Antoniadis I., 1999, arXiv:hep-th/9909212v1. 

\bibitem[\protect\citeauthoryear{Avelino, Martins, Nunes and Olive}{2006}]{avelino}
Avelino P. P., Martins C. J. A. P., Nunes N. J. and Olive K. A., 2006, Phys. Rev. D. 74, 083508.

\bibitem[\protect\citeauthoryear{Bahcall and Schmidt}{1967}]{bahcall}
Bahcall J. N. and Schmidt M., 1967, Phys. Rev. Lett. 19, 1294.

\bibitem[\protect\citeauthoryear{Bahcall, Sargent and Schmidt}{1967}]{bahcall1}
Bahcall J. N., Sargent W. and Schmidt M., 1967, Astrophys. J Lett. 149, L11.

\bibitem[\protect\citeauthoryear{Bak and Rey}{2000}]{bakrey}
Bak D. and Rey S. J., 2000, Class. Quant. Grav. 17, L83.

\bibitem[\protect\citeauthoryear{Banerjee and Sen}{1997}]{banerjeesen}
Banerjee N. and Sen S., 1997, Phys. Rev. D 56, 1334.

\bibitem[\protect\citeauthoryear{Banerjee, Cai, Heisenberg, Colgain, Sheikh-Jabbari and Yang}{2021}]{banerjee}
Banerjee A., Cai H., Heisenberg L., Colgain E. O., Sheikh-Jabbari M. M. and Yang T., 2021, Phys. Rev. D. 103, 081305.

\bibitem[\protect\citeauthoryear{Barker}{1978}]{barker}
Barker B. M., 1978, Astrophys. J. 219, 5.

\bibitem[\protect\citeauthoryear{Barrow}{1987}]{barrowsingle}
Barrow J. D., 1987, Phys. Rev. D. 35 , 1805.

\bibitem[\protect\citeauthoryear{Barrow}{1999}]{barrow1}
Barrow J. D., 1999, Phys. Rev. D. 59 , 043515.

\bibitem[\protect\citeauthoryear{Barrow and Magueijo}{1999}]{barrow2}
Barrow J. D. and Magueijo J., 1999, Phys. Lett. B. 447, 246.

\bibitem[\protect\citeauthoryear{Barrow and Magueijo}{1998}]{barrow3}
Barrow J. D. and Magueijo J., 1998, Phys. Lett. B. 443, 104.

\bibitem[\protect\citeauthoryear{Barrow, Sandvik and Magueijo}{2002}]{barrowprd}
Barrow J. D., Sandvik H. B. and Magueijo J., 2002, Phys. Rev. D. 65 : 063504.

\bibitem[\protect\citeauthoryear{Barrow, Magueijo and Sandvik}{2002}]{barrowplb}
Barrow J. D., Magueijo J. and Sandvik H. B., 2002, Phys. Lett. B. 541 : 201.

\bibitem[\protect\citeauthoryear{Barrow and Lip}{2012}]{barrowlip}
Barrow J. D. and Lip S. Z. W., 2012, Phys. Rev. D. 85, 023514.

\bibitem[\protect\citeauthoryear{Bekenstein}{1982}]{bekenstein}
Bekenstein J. D., 1982, Phys. Rev. D. 25, 1527.

\bibitem[\protect\citeauthoryear{Bernstein and Jain}{2004}]{bern}
Bernstein G. M. and Jain B., 2004, ApJ, 600, 17.

\bibitem[\protect\citeauthoryear{Bertotti, Iess and Tortota}{2003}]{bertotti}
Bertotti B., Iess  L. and Tortora P., 2003, Nature, 425, 374.

\bibitem[\protect\citeauthoryear{Betoule et al.}{2014}]{betoule}
Betoule M. et al., 2014, A and A, 568, A22.

\bibitem[\protect\citeauthoryear{Beutler et al.}{2011}]{beutler}
Beutler F. et al., 2011, MNRAS, 416, 3017.

\bibitem[\protect\citeauthoryear{Blake et. al.}{2012}]{blake}
Blake C. et al., 2012, MNRAS, 425, 405.

\bibitem[\protect\citeauthoryear{BOSS collaboration}{2012}]{boss}
BOSS collaboration, 2012, MNRAS, 441, 24.

\bibitem[\protect\citeauthoryear{Brans and Dicke}{1961}]{bransdicke}
Brans, C. and Dicke, R. H., 1961, Phys. Rev., 124, 925.

\bibitem[\protect\citeauthoryear{Brax et. al.}{2004}]{brax}
Brax P. et al., 2004, Phys. Rev. D 70, 123518.

\bibitem[\protect\citeauthoryear{Campbell and Olive}{1995}]{campbell}
Campbell B. A. and Olive K. A., 1995, Phys. Lett. B. 345 : 429.

\bibitem[\protect\citeauthoryear{Cattoen and Visser}{2007}]{catto}
Cattoen C. and Visser M., 2007, Class. Quant. Gravity, 24, 5985.

\bibitem[\protect\citeauthoryear{Chakrabarti}{2021}]{chakrabarti1}
Chakrabarti S., 2021, Mon. Not. Roy. Astron. Soc., 502 (2), 1895.

\bibitem[\protect\citeauthoryear{Chakrabarti}{2021}]{chakrabarti2}
Chakrabarti S., 2021, Mon. Not. Roy. Astron. Soc., 506, 2518.

\bibitem[\protect\citeauthoryear{Chand, Srianand, Petitjean and Aracil}{2004}]{chand}
Chand H., Srianand R., Petitjean P. and Aracil B., 2004, A and A, 417, 853.

\bibitem[\protect\citeauthoryear{Chiba}{2011}]{chiba}
Chiba T., 2011, Prog. Theor. Phys., 126, 993.

\bibitem[\protect\citeauthoryear{Chuang and Wang}{2013}]{chuang}
Chuang C. H. and Wang Y., 201, MNRAS, 435, 255.

\bibitem[\protect\citeauthoryear{Coc, Nunes, Olive, Uzan and Vangioni}{2007}]{coc}
Coc A., Nunes N. J., Olive K. A., Uzan J. and Vangioni E., 2007, Phys. Rev. D. 76 : 023511.

\bibitem[\protect\citeauthoryear{Copeland, Sami and Tsujikawa}{2006}]{copeland}
Copeland E. J., Sami M. and Tsujikawa S., 2006, Int. J. Mod. Phys. D, 15, 1753.

\bibitem[\protect\citeauthoryear{Cruz Perez and Sola}{2018}]{cruzperez}
Cruz Perez J., Sola J., 2018, Mod. Phys. Lett., A33, 1850228.

\bibitem[\protect\citeauthoryear{Damour and Esposito-Farese}{1992}]{damour1}
Damour, T. and Esposito-Farese, G., 1992, Class. Quant. Grav., 9, 2093.

\bibitem[\protect\citeauthoryear{Damour and Polyakov}{1994}]{damourpoly}
Damour T. and Polyakov A. M., 1994, Nucl. Phys. B 423, 532 ; Gen. Rel. Grav. 26, 1171.

\bibitem[\protect\citeauthoryear{Delubac et. al.}{2015}]{delubac}
Delubac T. et al., 2015, Astronomy and Astrophysics, 574, A59.

\bibitem[\protect\citeauthoryear{Bora and Desai}{2021}]{desai}
Bora K. and Desai S., 2021, JCAP, 02, 012.

\bibitem[\protect\citeauthoryear{Dicke}{1965}]{dicke}
Dicke R. H., 1965, The Theoretical Significance of Experimental Relativity (Gordon and Breach, New York).

\bibitem[\protect\citeauthoryear{Dirac}{1937}]{dirac1}
Dirac, P. A. M., 1937, Nature, 139, 323.

\bibitem[\protect\citeauthoryear{Dirac}{1938}]{dirac2}
Dirac, P. A. M., 1938, Proc. R. Soc. London, Ser. A, 165, 199.

\bibitem[\protect\citeauthoryear{Doran}{2005}]{doran}
Doran M., 2005, J. Cosmol. Astropart. Phys., 0504, 016.

\bibitem[\protect\citeauthoryear{Duff}{2014}]{duff1}
Duff, M. J., 2014, arXiv : 1412.2040v2 [hep-th], IMPERIAL-TP-2014-MJD-05.

\bibitem[\protect\citeauthoryear{Duff}{2016}]{duff2}
Duff, M. J., 2016, arXiv : hep-th/0208093v4.

\bibitem[\protect\citeauthoryear{Dunajski}{2008}]{duna}
Dunajski M. and Gibbons G., 2008, Class. Quant. Gravity, 25, 235012.

\bibitem[\protect\citeauthoryear{Dvali and Zaldarriaga}{2002}]{dvali}
Dvali G. R. and Zaldarriaga M., 2002, Phys. Rev. Lett. 88, 091303.

\bibitem[\protect\citeauthoryear{Dyson}{1967}]{dyson1}
Dyson F. J., 1967, Phys. Rev. Lett. 19, 1291.

\bibitem[\protect\citeauthoryear{Dyson}{1972}]{dyson2}
Dyson F. J., 1972, Aspects of Quantum Theory, edited by A. Salam and E. Wigner (Cambridge University Press,
London).

\bibitem[\protect\citeauthoryear{Eisenstein et al.}{2005}]{eisenstein}
Eisenstein D. J. et al., 2005, ApJ, 633, 560.

\bibitem[\protect\citeauthoryear{Evans et al.}{2014}]{evans}
Evans T. M., Murphy M. T., Whitmore J. B., Misawa T., Centurion M., D'Odorico S., Lopez S., Martins C. J. A. P., Molaro P., Petitjean P., Rahmani H., Srianand R. and Wendt M., 2014, MNRAS 445, 128.

\bibitem[\protect\citeauthoryear{Faraoni}{1999}]{faraoni}
Faraoni V., 1999, Phys. Rev. D 59, 084021.

\bibitem[\protect\citeauthoryear{Ferreira et. al.}{2014}]{ferreira1}
Ferreira M. C., Frigola O., Martins C. J. A. P., Monteiro A. M. R. V. L. and Sola J., 2014, Phys. Rev. D. 89, 083011.

\bibitem[\protect\citeauthoryear{Ferreira and Martins}{2015}]{ferreira2}
Ferreira M. C. and Martins C. J. A. P., 2015, Phys. Rev. D. 91, 124032.

\bibitem[\protect\citeauthoryear{Ferreira, Juliao, Martins and Monteiro}{2013}]{ferreira3}
Ferreira M. C., Juliao M. D., Martins C. J. A. P. and Monteiro A. M. R. V. L., 2013, Phys. Lett. B. 724, 1.

\bibitem[\protect\citeauthoryear{Ferreira, Juliao, Martins and Monteiro}{2012}]{ferreira4}
Ferreira M. C., Juliao M. D., Martins C. J. A. P. and Monteiro A. M. R. V. L., 2012, Phys. Rev. D. 86, 125025.

\bibitem[\protect\citeauthoryear{Fierz}{1956}]{fierz}
Fierz, M., 1956, Helv. Phys. Acta, 29, 128.

\bibitem[\protect\citeauthoryear{Frieman, Hill, Stebbins and Waga}{1995}]{frieman}
Frieman J. A., Hill C. T., Stebbins A. and Waga I., 1995, Phys. Rev. Lett. 75, 2077.

\bibitem[\protect\citeauthoryear{Foreman-Mackey, Hogg, Lang and Goodman}{2013}]{pythonmcmc}
Foreman-Mackey D., Hogg D. W., Lang D. and Goodman J., 2013, PASP, 125, 306.

\bibitem[\protect\citeauthoryear{Forgacs and Horvath}{1979}]{forgacs}
Forgacs P. and Horvath P., 1979, Gen. Relativ. Gravit. 11 , 205.

\bibitem[\protect\citeauthoryear{Frolov and Kofman}{2003}]{frolov}
Frolov A. V. and Kofman L., 2003, JCAP 0305, 009.

\bibitem[\protect\citeauthoryear{Gamow}{1967}]{gamow}
Gamow G., 1967, Phys. Rev. Lett. 19, 759.

\bibitem[\protect\citeauthoryear{Gibbons and Hawking}{1977}]{gibbons}
Gibbons G. W. and Hawking S. W., 1977, Phys. Rev. D. 15, 2738.

\bibitem[\protect\citeauthoryear{Gubser and Khoury}{2004}]{gubser}
Gubser S. S. and Khoury J., 2004, Phys. Rev. D 70, 104001.

\bibitem[\protect\citeauthoryear{Gupta}{2022}]{gupta}
Gupta, R. P., 2022, arXiv:2201.11667v1 [gr-qc].

\bibitem[\protect\citeauthoryear{Heisenberg, Villarrubia-Rojo and Zosso}{2022}]{heisenberg}
Heisenberg L., Villarrubia-Rojo H. and Zosso J., 2022, arXiv:2201.11623v1 [astro-ph.CO]. 

\bibitem[\protect\citeauthoryear{Hinterbichler and Khoury}{2010}]{hinter}
Hinterbichler K. and Khoury J., 2010, Phys. Rev. Lett. 104, 231301.

\bibitem[\protect\citeauthoryear{Jacobson}{1995}]{jacobson}
Jacobson T., 1995, Phys. Rev. Lett. 75, 1260.

\bibitem[\protect\citeauthoryear{Jain and Khoury}{2010}]{jain}
Jain B. and Khoury J., 2010, Annals Phys. 325, 1479.

\bibitem[\protect\citeauthoryear{Jamil, Saridakis and Setare}{2010}]{jamil}
Jamil M., Saridakis E. N. and Setare M. R., 2010, JCAP 1011, 032.

\bibitem[\protect\citeauthoryear{Jordan}{1937}]{jordan}
Jordan, P., 1937, Die Naturwissenschaften, 25, 513.

\bibitem[\protect\citeauthoryear{Khoury and Weltman}{2004}]{khoury}
Khoury J. and Weltman A., 2004, Phys. Rev. Lett. 93, 171104 ; Phys. Rev. D 69, 044026.

\bibitem[\protect\citeauthoryear{Kotus, Murphy and Carswell}{2017}]{kotus}
Kotus S. M., Murphy M. T. and Carswell R. F., 2017, MNRAS 464, 3679.

\bibitem[\protect\citeauthoryear{Landau, Sisterna and Vucetich}{2001}]{landau}
Landau S., Sisterna P. and Vucetich H., 2001, Phys. Rev. D. 63, 081303(R).

\bibitem[\protect\citeauthoryear{Lee, Lee, Colgain, Sheikh-Jabbari and Thakur}{2022}]{lee}
Lee B-H., Lee W., Colgáin E. O., Sheikh-Jabbari M. M. and Thakur S., 2022, arXiv:2202.03906v2 [astro-ph.CO]

\bibitem[\protect\citeauthoryear{Leite et. al.}{2014}]{leite}
Leite A. C. O., Martins C. J. A. P., Pedrosa P. O. J. and Nunes N. J., 2014, Phys. Rev. D. 90, 063519.

\bibitem[\protect\citeauthoryear{Lu et. al.}{2009}]{lu}
Lu J., et al., 2009, Int. J. Mod. Phys. D, 18, 1741.

\bibitem[\protect\citeauthoryear{Luo, Olive and Uzan}{2011}]{luo}
Luo F., Olive K. A. and J. Uzan, 2011, Phys. Rev. D. 84 : 096004.

\bibitem[\protect\citeauthoryear{Magueijo, Sandvik and Kibble}{2001}]{magukibble}
Magueijo J., Sandvik H. B. and Kibble T. W. B., 2001, Phys. Rev. D. 64 , 023521.

\bibitem[\protect\citeauthoryear{Maor and Brunstein}{2003}]{maor1}
Maor I. and Brustein R., 2003, Phys. Rev. D, 67, 103508.

\bibitem[\protect\citeauthoryear{Maor, Brunstein and Steinhardt}{2001}]{maor2}
Maor I., Brustein R. and Steinhardt P. J., 2001, Phys. Rev. Lett., 86, 6.

\bibitem[\protect\citeauthoryear{Marciano}{1984}]{marciano}
Marciano W., 1984, Phys. Rev. Lett. 52 , 489.

\bibitem[\protect\citeauthoryear{Martins}{2015}]{martins}
Martins C. J. A. P., 2015, Gen. Rel. Grav. 47, 1843.

\bibitem[\protect\citeauthoryear{Martins et. al.}{2015}]{martins1}
Martins C. J. A. P., Pinho A. M. M., Alves R. F. C., Pino M., Rocha C. I. S. A. and von Wietersheim M.,
2015, JCAP 1508, 047.

\bibitem[\protect\citeauthoryear{Martins and Pinho}{2017}]{martinspinho}
Martins C. J. A. P. and Pinho A. M. M., 2017, Phys. Rev. D. 95, 023008.

\bibitem[\protect\citeauthoryear{Moffat}{1993}]{moffat}
Moffat J., 1993, Int. J. Mod. Phys. D. 2, 351.

\bibitem[\protect\citeauthoryear{Molaro et. al.}{2013}]{molaro}
Molaro P., Centurion M., Whitmore J., Evans T., Murphy M. et al., 2013, A and A 555, A68.

\bibitem[\protect\citeauthoryear{Moresco et. al.}{2012}]{moresco}
Moresco M., Verde L., Pozzetti L., Jimenez R. and Cimatti A., 2012, J. Cosmol. Astropart. Phys, 07, 053.

\bibitem[\protect\citeauthoryear{Mukherjee and Banerjee}{2016}]{mukherjee}
Mukherjee A. and Banerjee N., 2016, Phys. Rev. D. 93, 043002.

\bibitem[\protect\citeauthoryear{Murphy, Webb and Flambaum}{2003}]{murphy}
Murphy M. T., Webb J. K. and Flambaum V. V., 2003, MNRAS, 345, 609.

\bibitem[\protect\citeauthoryear{Nordtvedt Jr.}{1970}]{nordt}
Nordtvedt Jr. K., 1970, Astrophys. J. 161, 1059.

\bibitem[\protect\citeauthoryear{Nunes and Lidsey}{2004}]{nunes}
Nunes N. J. and Lidsey J. E., 2004, Phys. Rev. D, 69, 123511.

\bibitem[\protect\citeauthoryear{Padmanabhan and Roychoudhury}{2003}]{paddy1}
Padmanabhan T. and Roychoudhury T., 2003, MNRAS, 344, 823.

\bibitem[\protect\citeauthoryear{Padmanabhan}{2003}]{paddythermo}
Padmanabhan T., 2003, Phys. Rept. 380, 235.

\bibitem[\protect\citeauthoryear{Roychoudhury and Padmanabhan}{2005}]{paddy2}
Roychoudhury T. and Padmanabhan T., 2005, Astron. Astrophys., 429, 807.

\bibitem[\protect\citeauthoryear{Parkinson, Bassett and Barrow}{2004}]{parkinson}
Parkinson D., Bassett B. A. and Barrow J. D., 2004, Phys. Lett. B, 578, 235.

\bibitem[\protect\citeauthoryear{Peres}{1967}]{peres}
Peres A., 1967, Phys. Rev. Lett. 19, 1293.

\bibitem[\protect\citeauthoryear{Pinho and Martins}{2016}]{pinho}
Pinho A. M. M. and Martins C. J. A. P., 2016, Phys. Lett. B. 756, 121.

\bibitem[\protect\citeauthoryear{Planck collaboration}{2014}]{planck}
Planck collaboration XVI, 2014, Astronomy and Astrophysics, 571, A16.

\bibitem[\protect\citeauthoryear{Reimers}{2012}]{reim}
Reimers D. and Kozlov M. G., 2012, A and A 540, L9.

\bibitem[\protect\citeauthoryear{Riess et. al.}{2001}]{riess1}
Riess A. G. et al., 2001, ApJ, 560, 49.

\bibitem[\protect\citeauthoryear{Riess et. al.}{2004}]{riess2}
Riess A. G. et al., 2004, ApJ, 607, 665.

\bibitem[\protect\citeauthoryear{Sahni and Starobinski}{2000}]{sahni}
Sahni V. and Starobinski A. A., 2000, Int. J. Mod. Phys. D, 9, 373.

\bibitem[\protect\citeauthoryear{Sahni, Saini, Starobinski and Alam}{2003}]{sahni1}
Sahni, V., Saini, T. D., Starobinsky A. A. and Alam, U., 2003, JETP Lett., 77, 201.

\bibitem[\protect\citeauthoryear{Sahni et. al.}{2008}]{sahni2}
Sahni, V., et al., 2008, Phys. Rev. D, 78,103502.

\bibitem[\protect\citeauthoryear{Sandvik, Barrow and Magueijo}{2002}]{bsbm}
Sandvik H. B., Barrow J. D., Magueijo J., 2002, Phys. Rev. Lett. 88, 031302.

\bibitem[\protect\citeauthoryear{Savedoff}{1956}]{savedoff}
Savedoff M. P., 1956, Nature 178, 689.

\bibitem[\protect\citeauthoryear{Schutzhold}{2002}]{ralf}
Schutzhold R., 2002, Int. J. Mod. Phys. A, 17(29), 4359. 

\bibitem[\protect\citeauthoryear{Schwinger}{1970}]{swinger}
Schwinger J., 1970, Particles, Sources and Fields (Addison-Wesley, Reading).

\bibitem[\protect\citeauthoryear{Sen, Sen and Sami}{2010}]{sensensami}
Sen S., Sen A. A. and Sami M., 2010, Phys. Lett. B, 686, 1

\bibitem[\protect\citeauthoryear{Shafieloo, Alam, Sahni and Starobinsky}{2006}]{shafieloo}
Shafieloo A., Alam U., Sahni V. and Starobinsky A. A., 2006, Mon. Not. Roy. Ast. Soc. 366, 1081.

\bibitem[\protect\citeauthoryear{Shlyakhter}{1976}]{akhter}
Shlyakhter A. I., 1976, Nature 264, 340.

\bibitem[\protect\citeauthoryear{Simon, Verde and Jimenez}{2005}]{jimenez}
Simon J., Verde L. and Jimenez R., 2005, Phys. Rev. D. 71, 123001.

\bibitem[\protect\citeauthoryear{Slepian, Gott and Zinn}{2014}]{slepian}
Slepian Z., Gott J. R. and Zinn J., 2014, MNRAS, 438, 1948.

\bibitem[\protect\citeauthoryear{Sola et. al.}{2019}]{sola1}
Sola J., Gomez-Valent A., Cruz Perez J., Moreno-Pulido C., 2019, ApJ, 886, L6.

\bibitem[\protect\citeauthoryear{Sola et. al.}{2020}]{sola2}
Sola J., Gomez-Valent A., Cruz Perez J., Moreno-Pulido C., 2020, Class. Quant. Grav., 37, 245003.

\bibitem[\protect\citeauthoryear{Sola, Karimkhani and Khodam-Mohammadi}{2017}]{sola3}
Sola J., Karimkhani E. and Khodam-Mohammadi A., 2017, Class. Quant. Grav. 34, no.2, 025006.

\bibitem[\protect\citeauthoryear{Songaila and Cowie}{2014}]{songaila}
Songaila A. and Cowie, L. 2014, Astrophys. J. 793, 103 (2014).

\bibitem[\protect\citeauthoryear{Stern et. al.}{2010}]{stern}
Stern D., Jimenez R., Verde L., Kamionkowski M. and Stanford S., 2010, J. Cosmol. Astropart. Phys, 02, 008.

\bibitem[\protect\citeauthoryear{Tong and Zhang}{2009}]{tong}
Tong M. L. and Zhang Y., 2009, Phys. Rev. D, 80, 023503.

\bibitem[\protect\citeauthoryear{Turneaure and Stein}{1976}]{turne}
Turneaure J. P. and Stein S. R., 1976, Atomic Masses and Fundamental Constants, edited by J. H.
Sanders and A. H. Wapstra (Plenum, New York).

\bibitem[\protect\citeauthoryear{Unzicker}{2009}]{unzicker}
Unzicker, A., 2009, Ann. Phys. (Berlin), 18(1), 57.

\bibitem[\protect\citeauthoryear{Upadhye, Ishak and Steinhardt}{2005}]{upadhye}
Upadhye A., Ishak M. and Steinhardt P. J., 2005, Phys. Rev. D, 72, 063501.

\bibitem[\protect\citeauthoryear{Upadhye, Gubser and Khoury}{2005}]{upadhye1}
Upadhye A., Gubser S. S. and Khoury J., 2006, Phys. Rev. D 74, 104024.

\bibitem[\protect\citeauthoryear{Uzan}{2003}]{uzan1}
Uzan J. P., 2003, Rev. Mod. Phys., 75, 403.

\bibitem[\protect\citeauthoryear{Uzan}{2011}]{uzan2}
Uzan J. P., 2011, Living Rev. Relativ., 14, 2.

\bibitem[\protect\citeauthoryear{Van den Bergh}{1982}]{vdb}
Van den Bergh N., 1982, Gen. Relativ. Gravity 14, 17.

\bibitem[\protect\citeauthoryear{Velten, Marttens and Zimdahl}{2014}]{velten}
Velten H. E. S., vom Marttens R. F. and Zimdahl W., 2014, Eur. Phys. J. C. 74(11), 3160.

\bibitem[\protect\citeauthoryear{Visser}{2005}]{visser}
Visser M., 2005, Gen. Rel. Grav., 37, 1541.

\bibitem[\protect\citeauthoryear{Wang, Hui and Khoury}{2012}]{wang}
Wang J., Hui L. and Khoury J., 2012, Phys. Rev. Lett. 109, 241301.

\bibitem[\protect\citeauthoryear{Wang and Tegmark}{2005}]{wangteg}
Wang Y. and Tegmark M., 2005, Phys. Rev. D 71, 103513.

\bibitem[\protect\citeauthoryear{Webb et. al.}{2001}]{webbetal}
Webb J. K., Murphy M. T., Flambaum V. V., Dzuba V. A., Barrow J. D., Churchill C. W., Prochaska J. X. and Wolfe A. M., 2001, Phys. Rev. Lett., 87, 091301.

\bibitem[\protect\citeauthoryear{Webb et. al.}{2011}]{webb1}
Webb J. K., King J. A., Murphy M. T., Flambaum V. V., Carswell R. F. et al., 2011, Phys. Rev. Lett. 107, 191101.

\bibitem[\protect\citeauthoryear{Weinberg}{1972}]{weinberg}
Weinberg S., 1972 \textit{Gravitation and Cosmology} (New York: Wiley).

\bibitem[\protect\citeauthoryear{Whitmore and Murphy}{2015}]{whitmore}
Whitmore J. B. and Murphy M. T., 2015, Mon. Not. Roy. Astron. Soc. 447, 446.

\bibitem[\protect\citeauthoryear{Will}{2001}]{will0}
Will C. M., 2001, Liv. Rev. Rel. 4, 4.

\bibitem[\protect\citeauthoryear{Will}{2005}]{will}
Will C. M., 2005, Liv. Rev. Rel. 9, 3.

\bibitem[\protect\citeauthoryear{Wolfe, Brown and Roberts}{1976}]{wolfe}
Wolfe A. M., Brown, R. L. and Roberts M. S., 1976, Phys. Rev. Lett. 37, 179.

\bibitem[\protect\citeauthoryear{Zlatev, Wang and Steinhardt}{1999}]{zlatev}
Zlatev I., Wang L. and Steinhardt P. J., 1999, Phys. Rev. Lett. 82, 896 ; Phys. Rev. D. 59, 12350.








\end{thebibliography}

\end{document}